\documentclass[12pt]{article}
\usepackage{amsmath,amsfonts, amssymb}
\usepackage{hyperref}
\usepackage{graphicx}
\usepackage{psfrag}

\newcommand{\bea}{\begin{eqnarray}}
\newcommand{\eea}{\end{eqnarray}}
\newcommand{\ra}{\rightarrow}

\newcommand{\be}{\begin{equation}}
\newcommand{\ee}{\end{equation}}
\newcommand{\ba}{\begin{eqnarray}}
\newcommand{\ea}{\end{eqnarray}}
\newcommand{\bi}{\begin{itemize}}
\newcommand{\ei}{\end{itemize}}
\newcommand{\tr}{{\rm tr}}
\newcommand{\Tr}{{\rm Tr}}

\newcommand{\Z}{{\mathbb Z}}
\newcommand{\R}{\mathbb{R}}

\newcommand{\Ncal}{{\mathcal N}}

\newcommand{\Wcal}{{\mathcal W}}

\newcommand{\gfrak}{{\mathfrak g}}

\newcommand{\tfrak}{{\mathfrak t}}
\newcommand{\ufrak}{{\mathfrak u}}
\newcommand{\nn}{\nonumber}
\newcommand{\mo}{{-1}} 

\newcommand{\f}{\frac}
\newcommand{\half}{\frac{1}{2}}
\newcommand{\oo}{\frac{1}}

\newcommand{\aslash}[1]{\,\,{\raise.15ex\hbox{/}\mkern-12mu #1}}
\newcommand{\bslash}[1]{\,\,{\raise.15ex\hbox{/}\mkern-9mu #1}}

\newcommand{\LR}{{}^L\negthinspace R}
\newcommand{\LG}{{}^L\negthinspace\hspace{.4mm} G}
\newcommand{\Lv}{{}^L\negthinspace\hspace{.4mm} v}
\newcommand{\LLv}{\raisebox{.8mm}{\mbox{\it \scriptsize{L}}} \mspace{1mu} v }
\newcommand{\Lw}{{}^L\negthinspace\hspace{.4mm} w}
\newcommand{\LLw}{\raisebox{.9mm}{\mbox{\it \scriptsize{L}}} \mspace{1mu} w }
\newcommand{\Lalpha}{{}^L\negthinspace\hspace{.4mm} \alpha}
\newcommand{\LLalpha}{\raisebox{1mm}{\mbox{\it \scriptsize{L}}}\alpha }

\newcommand{\LH}{{}^L\negthinspace H}

\newcommand{\LL}{{}^L\negthinspace}

\newcommand{\LLt}{\raisebox{1mm}{\mbox{\it \scriptsize{L}}}\mspace{.3mu}\mathfrak t}
\newcommand{\Ltau}{{}^L\negthinspace\hspace{.4mm}  \tau}
\newcommand{\LLtau}{\raisebox{.8mm}{\mbox{\it \scriptsize{L}}}\mspace{.3mu}\tau}

\newcommand{\LLgc}{\raisebox{.8mm}{\mbox{\it \scriptsize{L}}} g }
\newcommand{\Ltheta}{{}^L\negthinspace\hspace{.4mm} \theta}
\newcommand{\Lgfrak}{{}^L\negthinspace\hspace{.4mm} {\mathfrak g}}
\renewcommand{\bar}{\overline}

\renewcommand{\hat}{\widehat}

\renewcommand{\title}[1]{\vbox{\center\LARGE{#1}}\vspace{5mm}}
\renewcommand{\author}[1]{\vbox{\center#1}\vspace{5mm}}
\newcommand{\address}[1]{\vbox{\center\em#1}}

\begin{document}
\bibliographystyle{utphys}

\begin{titlepage}
\hspace{11cm} {\tt NSF-KITP-09-58}\\
\begin{center}
\rightline{ }
\centerline{\LARGE Quantum 't Hooft operators and $S$-duality} \medskip\medskip
\centerline{\LARGE in ${\cal N}=4$ super Yang-Mills }
\vskip 10mm
Jaume Gomis${}^{a,}$\footnote{{\rm jgomis(at)perimeterinstitute.ca}
}, Takuya Okuda${}^{a,b,}$\footnote{\rm takuya(at)perimeterinstitute.ca}
 and
Diego Trancanelli${}^{b,c,}$\footnote{\rm dtrancan(at)physics.ucsb.edu}

\vskip 5mm

\address{
${}^a$Perimeter Institute for Theoretical Physics,

Waterloo, Ontario, N2L 2Y5, Canada
}
\address{
${}^b$Kavli Institute for Theoretical Physics, University of California,

Santa Barbara, CA 93106, USA
}
\address{
${}^c$Physics Department,
University of California,

Santa Barbara, CA 93106, USA
}
\vskip 3mm
\vskip 3mm

\end{center}

\abstract{

\noindent \normalsize{We provide a quantum path integral definition
of an 't Hooft loop operator, which inserts a pointlike monopole in
a four-dimensional gauge theory. We explicitly compute the
expectation value of the circular  't Hooft operators in $\Ncal=4$
super Yang-Mills with arbitrary gauge group $G$ up to next to
leading order in   perturbation theory. We  also compute in the
strong coupling expansion   the expectation value of the  circular
Wilson loop operators. The result of the  computation of an 't Hooft
loop operator in the weak coupling expansion exactly reproduces the
strong coupling result of the conjectured dual Wilson loop operator
under the action of $S$-duality. This paper demonstrates -- for the
first time -- that  correlation functions in  $\Ncal=4$ super
Yang-Mills admit the action of $S$-duality.

 }
 }

\vfill

\end{titlepage}

\tableofcontents


\setcounter{footnote}{0}

\section{Introduction}

Electric-magnetic duality, also known as $S$-duality
\cite{Montonen:1977sn, Witten:1978ma,Osborn:1979tq},  is a
remarkable conjectured equivalence relating   ${\cal N}=4$ super
Yang-Mills at weak coupling to ${\cal N}=4$ super Yang-Mills at
strong coupling. Heuristically, this equivalence arises via a change
of variables in the path integral,  which identifies the two
descriptions. This kind of   duality transformation can be
explicitly performed in certain statistical mechanics models such as
the Ising model \cite{Kramers:1941kn} as well as in
electromagnetism, where electric fields are replaced by magnetic
fields.  $S$-duality in ${\cal N}=4$ super Yang-Mills conjecturally
extends the  electric-magnetic duality transformation in
electromagnetism \cite{Seiberg:1994rs,Witten:1995gf}  to a
full-fledged interacting quantum field theory.

$S$-duality conjectures that ${\cal N}=4$ super Yang-Mills with
gauge group $G$ and coupling constant $\tau$ is equivalent to ${\cal
N}=4$ super Yang-Mills with dual gauge group  $\LG$
\cite{Goddard:1976qe} and coupling constant $\LLtau$. The coupling
constants of the two theories are related by
\ba
\LLtau=-\f{1}{n_\gfrak \tau}\,, \nonumber
\ea
where
\ba \tau={\theta\over 2\pi}+{4\pi i\over g^2}, \qquad
\LLtau={\Ltheta\over 2\pi}+{4\pi i\over (\LLgc)^2}\,,
\nonumber
\ea
and $n_\gfrak=1,2$ or $3$ depending\footnote{\label{ng} Here
$n_\gfrak=1$ for simply laced algebras; $n_\gfrak=2$ for
$\mathfrak{so}(2N+1) ,\mathfrak{sp}(N)$ and $\mathfrak{f}_4$; and
$n_\gfrak=3$ for $\mathfrak{g}_2$.} on the choice of gauge group
$G$. $S$-duality also acts on  all gauge invariant operators  of the
theory and defines an operator isomorphism between the two theories
\ba {\cal O}\longleftrightarrow {}^L\negthinspace\hspace{.4mm} {\cal
O}\,. \nonumber
\ea
Even though this map is rather poorly understood,  progress   in
recent years has resulted in conjectures relating a large class of
supersymmetric operators supported on various submanifolds in
spacetime.

Since $S$-duality interchanges electric and magnetic charges, it
exchanges a Wilson operator \cite{Wilson:1974sk} with an 't Hooft
operator \cite{'tHooft:1977hy}. These operators insert an
electrically charged source and a magnetically charged source,
respectively. Whereas a Wilson operator in the theory with gauge
group $G$ is labeled by a representation $R$ of $G$, an 't Hooft
operator  is labeled \cite{Kapustin:2005py} by a representation
$\LR$ of the dual group $\LG$, and will be denoted by  $W(R)$,
$T(\LR)$ respectively. Therefore, it is conjectured that under
$S$-duality \cite{Kapustin:2005py}
\ba
T(\LR)\longleftrightarrow W(\LR)\,. \nonumber
\ea
Explicit conjectures have also been made for the action of
$S$-duality on chiral primary operators \cite{Intriligator:1998ig,
Intriligator:1999ff, Argyres:2006qr}, surface operators
\cite{Gukov:2006jk, Gomis:2007fi} and domain walls
\cite{Gaiotto:2008sa, Gaiotto:2008ak} in ${\cal N}=4$ super
Yang-Mills.

The $S$-duality conjecture goes beyond the mapping of   operators.
It also predicts that the correlation functions of gauge invariant
operators -- which span the set of observables in ${\cal N}=4$ super
Yang-Mills -- are related in the two theories by
\ba
\bigg\langle \prod_i {\cal
O}_i\bigg\rangle_{G,\tau}=\bigg\langle \prod_i
{}^L\negthinspace\hspace{.4mm} {\cal O}_i\bigg\rangle_{\LG,\Ltau}\,.
\nonumber
\ea
This aspect of the $S$-duality conjecture  is a particularly
challenging one  to exhibit, as proving it necessarily requires
understanding correlation functions at strong coupling, where no
universal methods of computation are readily available.

In this paper we exhibit -- for the first time -- that the
correlation function of dual operators are mapped into each other
under the action of $S$-duality. We show that the weak coupling
computation of the circular 't Hooft operator $T(\LR)$ in ${\cal
N}=4$ super Yang-Mills with gauge group $G$  exactly reproduces the
strong coupling computation of  the expectation value of the
circular Wilson operator $W(\LR)$ in ${\cal N}=4$ super Yang-Mills
with gauge group $\LG$. We explicitly show that the prediction of
$S$-duality
\ba \langle T(\LR)\rangle_{G,\tau}= \langle
W(\LR)\rangle_{\LG,\Ltau}\, \label{prediccion}
\ea
holds to next to leading order in the coupling constant expansion,
which is weak for the 't Hooft operator and strong for the dual
Wilson operator.

Our computations verify in a quantitative manner the main prediction
of $S$-duality  for this  class of observables.  These results go
beyond the previous tests of $S$-duality, which involve quantities
for which the semiclassical approximation is exact or the theory is
topologically twisted. Such tests include  comparing  the BPS
spectra of particles \cite{Sen:1994yi} and operators
\cite{Kapustin:2005py}-\cite{Gaiotto:2008ak}, the effective action
\cite{Sen:1994fa}  in the Coulomb branch, and the partition function
of the theory \cite{Vafa:1994tf}. We note that the Wilson  and 't
Hooft operators that we consider in this paper are different than
the corresponding operators considered by Kapustin  and Witten
\cite{Kapustin:2006pk} in the topologically twisted ${\cal N}=4$
super Yang-Mills theory relevant for the gauge theory approach to
the geometric Langlands program. The Wilson  and 't Hooft operators
considered in that theory are for arbitrary curves and  have trivial
expectation values.

Exhibiting $S$-duality for Wilson and 't Hooft operators first
requires defining and computing the expectation value of an 't Hooft
operator  in ${\cal N}=4$ super Yang-Mills. In Section $2$ we
provide a quantum  definition of an 't Hooft operator   in
four-dimensional gauge theory. It is defined in terms of a path
integral where we integrate over all fields which have a prescribed
singularity near the operator. Properly defining the 't Hooft
operator $T(\LR)$ requires both renormalizing the operator as well
as completely specifying the measure of integration in the path
integral. The classical singularity that  we quantize is that of a
singular monopole,  which  is characterized \cite{Kapustin:2005py}
by the highest weight   $B$ of the representation $\LR$ under which
the 't Hooft operator $T(\LR)$ transforms. Demanding that the path
integral definition of the 't Hooft operator $T(\LR)$ is gauge
invariant requires  integrating over the $G$-orbit of the classical
singularity, which depends on  $B$, and results in the inclusion of
the measure of  the adjoint orbit of $B$ in the path integral
measure. This quantum prescription applies to the computation of
a general 't Hooft operator in an arbitrary gauge theory.
We explicitly compute the expectation value of the circular
't Hooft operator  $T(\LR)$  in ${\cal N}=4$ super Yang-Mills with
arbitrary gauge group $G$ up to one loop order. Given the path
integral definition we provide, the computation of the expectation
value of the 't Hooft operator $T(\LR)$  can be extended to higher
orders in perturbation theory by summing over the connected vacuum
diagrams generated by the path integral.

In Section $3$ we compute at strong coupling the expectation value
of the circular Wilson loop operator $W(R)$ in ${\cal N}=4$ super
Yang-Mills with arbitrary gauge group $G$. It was conjectured in
\cite{Erickson:2000af,Drukker:2000rr} that the expectation value of
the circular Wilson loop with $U(N)$ gauge group can be computed
using  a  Gaussian matrix model, thereby reducing the complexity of
the path integral of a four-dimensional field theory to a matrix
integral. This result, extended to  an arbitrary gauge group $G$ has
been proven by Pestun \cite{Pestun:2007rz}, who, using localization
techniques, has shown that the path integral over the
four-dimensional fields reduces to an integral over a zero mode,
which  corresponds to the variable of integration in the matrix
model integral. We use this result  to evaluate the Wilson loop
expectation value at strong coupling by performing the strong
coupling expansion of the corresponding matrix integral.

In Section $4$ we use the results of our computations of the 't
Hooft operator at weak coupling and of the dual Wilson   operator at
strong coupling and explicitly show that these correlators transform
precisely as conjectured by $S$-duality, and indeed verify equation
(\ref{prediccion}). In Section $5$ we argue that the subleading
exponential corrections that appear in the Wilson loop computation
can also be understood from the perturbative computation of the 't
Hooft operator around extra saddle points.  These saddle points
arise due to the physics of monopole screening, whereby the charge
of an 't Hooft operator is reduced/screened when a regular monopole
configuration approaches the operator. Inclusion of these saddle
points in the computation of the 't Hooft operator exactly
reproduces the strong coupling result for the Wilson loop operator.
Section $6$ contains a summary and discussion of our results and
future lines of inquiry. We have relegated to the appendices the
details of  some of our computations.


\section{'t Hooft loop expectation value}
\label{sec-thooft}

In this section we provide a quantum path integral definition of an
't Hooft operator in four-dimensional gauge theory, and explicitly
compute the expectation value of the circular 't Hooft loop operator
in ${\cal N}=4$ super Yang-Mills up to one loop order. We begin by
introducing basic facts regarding the classical field configuration
produced by an 't Hooft operator and then proceed to its
quantization.

't Hooft originally defined \cite{'tHooft:1977hy} these operators by
specifying a singular gauge transformation around an arbitrary curve
that links the loop on which the 't Hooft operator is
supported.\footnote{\label{spacelike} For a space-like curve, such a
singular gauge transformation    creates a magnetic flux tube  along
the loop. Thus an 't Hooft/Wilson loop can be interpreted as the
operator that creates an infinitesimally thin magnetic/electric flux
tube around the loop (see {\it e.g.} \cite{Alford:1992yx}).}
Therefore,  in a gauge theory with gauge group $G$, these operators
are labeled by $\pi_1(G)$, which measures the topological magnetic
flux created by the operator.

Kapustin \cite{Kapustin:2005py} --  motivated by $S$-duality in
${\cal N}=4$ super Yang-Mills --  has further refined 't Hooft's
original characterization of magnetic   operators and has shown that
't Hooft operators\footnote{We will  name this broader class of
operators also as 't Hooft operators.}  in a gauge theory with gauge
group $G$  are labeled by a representation  $\LR$ of the dual group
$\LG$. Since $\pi_1(G)\simeq Z(\LG)$, where $Z(\LG)$ is the center
of $\LG$, the topological magnetic flux  created by an operator
labeled by a representation $\LR$   is given by the charge
$Z(\LG)\subset \LG$ of the representation $\LR$ of
$\LG$.\footnote{This charge is the conjugacy class of the
representation, {\it i.e.}, the highest weight  modulo elements of
the root lattice, which coincides with  $N$-ality for $SU(N)$.}
Kapustin's classification is much finer, as  there are (infinitely)
many different operators for a given topological flux in $\pi_1(G)$.

Physically, an 't Hooft loop operator is an operator that inserts a
probe point-like monopole  whose worldline forms the loop in
spacetime on which the 't Hooft operator is supported.  The
representation  $\LR$ of $\LG$ which labels the operator
characterizes the magnetic charge of the  monopole
\cite{Goddard:1976qe}. This description parallels the more familiar
discussion of a Wilson loop operator, which inserts a point-like
electric charge, and is therefore labeled by a representation $R$ of
$G$. Unlike a Wilson operator, which can be described by the
insertion of an operator made out of the fields appearing in the
Lagrangian, an 't Hooft operator is defined by specifying a
singularity along the loop for the microscopic fields that we
integrate over in the path integral, and is therefore an example of
 a disorder operator \cite{Kadanoff:1970kz}.

The classical field configuration produced by an 't Hooft loop
operator $T(\LR)$ supported on an arbitrary curve $C\subset \R^4$ is
obtained by specifying a singularity for the fields near each point
in the loop. Near each point in the loop $C$,  the local singularity
is that associated to   a straight line  $\R$,  and the
singularities  created by an 't Hooft loop operator supported on a
general curve $C$ can be constructed by patching together the local
singularities  for the 't Hooft operator supported on a  straight
line $\R$. In a given theory,  an 't Hooft operator creates a
codimension three singularity for the fields that appear in the
classical action. The only restriction on the admissible codimension
three singularities created by an 't Hooft operator  is that they
solve the equations of motion of the theory  in $\R^4\setminus C$.

In the rest of the paper we focus our attention on ${\cal N}=4$
super Yang-Mills with gauge group $G$. The locally supersymmetric
singularity created by an 't Hooft operator $T(\LR)$ supported on a
straight line $\R\subset \R^4$  and labeled by a representation
$\LR$ of $\LG$ is given by \cite{Kapustin:2005py}
\ba
F={B\over 2}\, \hbox{vol}(S^2)+i g^2\theta {B\over
16\pi^2}{dt\wedge dr\over r^2}\,, \qquad \phi=\f{B}{2r}{g^2\over
4\pi} |\tau|\,. \label{def-thooft}
\ea
The straight line $\R$ is spanned by the coordinate $t$, $r$ is the
distance from the line and ${\rm vol}(S^2)$ is the volume form on
the two-sphere that surrounds the line $\R$. $B\equiv B^i {H_i} \in
\tfrak$ takes values in the Cartan subalgebra of the Lie algebra
$\gfrak$ associated with the gauge group $G$. As shown in
\cite{Goddard:1976qe}, the Dirac quantization condition $\exp(2\pi i
B)=id_G$ implies that  $B$ can be identified with the highest weight
of the  representation $\LR$ of the dual group $\LG$, justifying the
labeling of 't Hooft operators in terms of representations of the
dual group \cite{Kapustin:2005py}. The 't Hooft operator creates a
magnetic field through the $S^2$ surrounding the monopole,  and when
$\theta\neq 0$ it also generates an electric field, as the monopole
acquires electric charge via the Witten effect \cite{Witten:1979ey}.
Unbroken supersymmetry at a point in the loop requires that a scalar
field $\phi \equiv n^I\phi^I$ in the ${\cal N}=4$ super Yang-Mills
multiplet (here $(n^I)$ is a unit vector in $\R^6$) acquires a pole
near the loop with fixed residue.

We now consider  't Hooft operators that preserve maximal
supersymmetry. Preservation of sixteen supercharges  everywhere in
the loop $C$ requires that the 't Hooft loop is supported on two
possible curves --   $C=\R$ or $C=S^1$ -- which are related by a
global conformal transformation. The symmetry preserved by the
straight and circular 't Hooft operators   is  $OSp(4^*|4)\subset
PSU(2,2|4)$.\footnote{This is the supergroup for a maximally
supersymmetric 't Hooft loop in $\R^{1,3}$. Supersymmetric 't Hooft
loops exist on both $\R^{1,3}$ and $\R^4$. The corresponding
symmetry group for the dual Wilson loop was exhibited in
\cite{Bianchi:2002gz}.} The bosonic subgroup is $SO(4^*)\times
USp(4)$, where $SO(4^*)\simeq SU(1,1)\times SU(2)$ is the subgroup
of the four-dimensional conformal group $SU(2,2)$ preserving the
curve $\R$ or $S^1\subset \R^4$ and $USp(4)\simeq SO(5)\subset
SU(4)$ is left unbroken by the choice of the scalar field which
develops a pole near the loop.

We are now ready to proceed with the quantum definition of the 't
Hooft operator. The 't Hooft loop expectation value is specified by
a path integral where one  integrates over all fields which have the
prescribed singularity (\ref{def-thooft}) along the loop. In order
to give a complete definition of the operator, the precise measure
of integration needs to be determined. Before proceeding with the
study of the measure, we first   analyze     the leading
semiclassical result for the 't Hooft loop expectation value.


\subsection{Semiclassical 't Hooft Loop}

The semiclassical evaluation of the path integral requires expanding
the ${\cal N}=4$ super Yang-Mills path integral around the monopole
singularity
\ba A&=A_0+\hat{A}\cr \phi^I&=\phi_0^I+\hat{\phi}^I\,, \nonumber \ea
where $(A_0,\phi_0^I)$ is the classical singularity
(\ref{def-thooft}) corresponding to an 't Hooft operator $T(\LR)$
and $(\hat{A},\hat{\phi}^I)$ are the non-singular quantum
fluctuations that we must integrate over in the path integral.

The ${\cal N}=4$ super Yang-Mills action can be obtained by
dimensional reduction  of the ten-dimensional ${\cal N}=1$ super
Yang-Mills with the inclusion of the topological term\footnote{Here
and throughout  it is understood that derivatives with respect to
$M=4,5,\ldots, 9$ are trivial.}
\ba S= \oo{g^2}\int d^4x \sqrt h\, \tr\left[ \half   F^{MN}  F_{MN}
+i   \bar \psi \Gamma^M D_M \psi\right]  -i{\theta\over
8\pi^2}\int\, \tr\left(F\wedge F\right)\,,
\label{10Daction} \ea
where $\tr(\ ,\ )$ is the invariant metric on the Lie algebra
$\gfrak$ associated with the gauge group $G$ and $(A_M,\psi)$ are
the ten-dimensional gauge field and gaugino respectively. The metric
on the Lie algebra is normalized so that the short coroots of
$\gfrak$ have length-squared equal to  two. In this normalization
the topological term equals $ i\theta$ for the minimal instanton,
$\theta$ has period $2\pi$ and the complexified coupling constant is
given by \ba \tau={\theta\over 2\pi}+{4\pi i \over g^2}\,. \nonumber
\ea In terms of the four-dimensional fields in the ${\cal N}=4$
super Yang-Mills multiplet $A_M=(A_\mu,\phi^I)$, where
$\mu=0,\ldots,3$ and $I=4,\ldots 9$, the non-topological part of the
action reads\footnote{This expression is valid on an arbitrary
curved background with metric $h$ as long as we add to the action
the conformal coupling of the scalars $\sqrt{h}R\,
\tr(\phi^I\phi^I)/6$.}
\ba
\oo{g^2}\int d^4x \sqrt{h}\, \tr\left[
\half   F^{\mu\nu}  F_{\mu\nu} +D^\mu\phi^ID_\mu\phi^I+\half
[\phi^I,\phi^J]^2+i   \bar \psi \Gamma^\mu D_\mu \psi +\bar \psi
\Gamma^I  [\phi^I,\psi] \right]\,.
\nonumber
\ea

In the leading semiclassical approximation the expectation value of the 't Hooft operator $T(\LR)$ is given by
\ba
\langle T(\LR) \rangle_{G,\tau}\simeq
\exp\left(-S_{(0)}\right)\,, \label{onshell}
\ea
where $S_{(0)}$ is the ${\cal N}=4$ super Yang-Mills action
(\ref{10Daction}) evaluated on the classical singularity
(\ref{def-thooft}) created by the 't Hooft operator $T(\LR)$. In the
leading semiclassical approximation the quantum fluctuations
$(\hat{A},\hat{\phi}^I)$ are neglected.

In order to analyze the 't Hooft   operators $T(\LR)$ supported on
$C=\R$ and $C=S^1$ it is instructive to consider   ${\cal N}=4$
super Yang-Mills  in $AdS_2\times S^2$ instead of $\R^4$. As already
mentioned, these operators preserve an $SU(1,1)\times SU(2)$
subgroup of the four-dimensional conformal group, and in
$AdS_2\times S^2$ these symmetries are manifest, since they act as
isometries, while in $\R^4$ they act as conformal symmetries. We can
go between $\R^4$ and $AdS_2\times S^2$ by performing a Weyl
transformation
\ba
ds_{\R^4}^2=\Omega^2 ds^2_{AdS_2\times S^2}\,,
\nonumber
\ea
which is a classical symmetry of ${\cal N}=4$ super Yang-Mills. When
considering the 't Hooft operator supported on $C=\R$ the metric on
$AdS_2$ is  the upper half-plane metric while when the operator is
supported on $C=S^1$ the metric on $AdS_2$  is the metric on the
Poincar\'e disk (see Appendix   \ref{Weyl} for the explicit Weyl
transformations). For both choices of curve $C$, the 't Hooft
operator is supported at the conformal boundary of $AdS_2\times
S^2$, which is $C=\R$ for the upper half-plane metric and $C=S^1$
for the metric on the Poincar\'e disk.

 Insertion of an 't Hooft loop operator  $T(\LR)$ at the conformal boundary of $AdS_2\times S^2$ creates the following field configuration
\ba
F= \f{B}2 {\rm vol}(S^2)+i g^2\theta {B\over 16\pi^2}
{\rm vol}(AdS_2),~~~\phi=\f{B}{2}{g^2\over 4\pi} |\tau|\,.
\label{theta-background}
\ea
Since the $S^2$ is non-contractible and the scalar field is
homogeneous in $AdS_2\times S^2$,  the field configuration created
by the  't Hooft operator  $T(\LR)$  in $AdS_2\times S^2$ is
non-singular.

We can now calculate the expectation value of the 't Hooft operator
$T(\LR)$ by evaluating the ${\cal N}=4$ super Yang-Mills action
(\ref{10Daction}) in $AdS_2\times S^2$ on the  field configuration
in equation (\ref{theta-background}) produced by the 't Hooft
operator  $T(\LR)$. In the action, $h$ refers to the  metric in
$AdS_2\times S^2$.\footnote{Since the scalar curvature on
$AdS_2\times S^2$ vanishes, the  conformal coupling vanishes.} Since
the scalar field is homogeneous, the on-shell ${\cal N}=4$ super
Yang-Mills  action is given by
\ba
S_{(0)}={1\over g^2} \int \tr(F_0\wedge *F_0)-i{\theta\over
8\pi^2} \int \tr(F_0\wedge F_0)=\tr(B^2){g^2|\tau|^2\over 16\pi}{\rm
Vol}(AdS_2)\,. \label{bulkaction}
\ea
The on-shell action is divergent, being proportional to the volume
of $AdS_2$. This result is as expected, since the on-shell action
measures the energy of an infinitely heavy pointlike magnetic
monopole.

In quantum field theory, the observables that are finite are the
correlation functions of renormalized operators.  Therefore, we must
appropriately renormalize the 't Hooft operator $T(\LR)$, which we
do as follows.  We first parametrize the metric near the boundary of
$AdS_2$ using the Fefferman-Graham gauge
\ba
ds^2_{AdS_2}={dZ^2\over Z^2}+{dX^2\over Z^2}\left(g_{0}(X)+Z^2
g_{2}(X)+\ldots\right)\,. \nonumber
\ea
In this coordinate system the boundary is at $Z=0$, and $X$
parametrizes $\R$ or $S^1$ for the upper half-plane metric and
Poincar\'e disk metric respectively.\footnote{$Z=2e^{-\rho},
X=\psi$ for the Poincar\'e disk, and $Z=l$, $X=t$ for the upper
half-plane (see equations (\ref{ads2s2}) and (\ref{UHP}) in Appendix
\ref{Weyl}). } In order to define the renormalized 't Hooft operator
we introduce a cutoff near the location of the operator, which is
inserted at the boundary of $AdS_2\times S^2$.\footnote{The definition of
the 't Hooft operator as  
the partition function of  ${\cal N}=4$ super Yang-Mills  on $AdS_2\times S^2$
is reminiscent of Sen's   definition of the quantum entropy
function \cite{Sen:2008yk, Gupta:2008ki, Sen:2008vm, Sen:2009vz} as the 
string theory  path  integral  on $AdS_2$, which  encodes  the macroscopic degeneracy of states of extremal
black holes.}  This defines a
three-dimensional hypersurface $\Sigma$ located at $Z=\epsilon$. The
renormalized 't Hooft operator is constructed by adding to the
${\cal N}=4$ super Yang-Mills action (\ref{10Daction}) covariant
counterterms supported on the hypersurface $\Sigma$
\ba
S\longrightarrow S+S_{ct}\,.
\nonumber
\ea
The explicit form of the covariant counterterms we use to define the
renormalized 't Hooft operator are the boundary terms\footnote{The
boundary terms for surface operators \cite{Gukov:2006jk} (see also
\cite{Gomis:2007fi}) were constructed in \cite{Drukker:2008wr}.}
\ba
S_{ct}=-\f{1}{g^2} \int_\Sigma\, \tr \left[ F|_\Sigma \wedge
\star_3 F|_\Sigma  - f\wedge \star_3  f\right]\,,
  \label{counterterm2}
\ea
where $F|_\Sigma$ is the restriction of $F$ to the hypersurface
$\Sigma$, $f$ is a one-form obtained by contracting $F$ with the
unit normal vector to   $\Sigma$ and $\star_3$ is the Hodge star
operation on the three-dimensional hypersurface.

Taking into account the bulk action (\ref{bulkaction}) and the
boundary terms (\ref{counterterm2}) in the semiclassical evaluation
of the expectation value of the circular 't Hooft operator  $T(\LR)$
we obtain that\footnote{ The net effect of the boundary terms is to
renormalize the volume of $AdS_2$. For the metric on the upper half
plane the renormalized volume vanishes while the renormalized volume
in the Poincar\'e disk is $-2\pi$, a well known result from studies
of Wilson loops in the AdS/CFT correspondence.}
\ba
\langle T(\LR) \rangle_{G,\tau}= \exp\left({\tr(B^2) \over 8}
g^2|\tau|^2\right)\,. \label{circleresult}
\ea
When the 't Hooft loop is supported on $C=\R$ the expectation value
is trivial, a result that follows from supersymmetry.

Exactly the same results for the semiclassical expectation value of
the 't Hooft operator $T(\LR)$ are obtained  when we consider the
theory   on $\R^4$. Everything we have done can be translated into
the  $\R^4$ language by performing a Weyl
transformation.\footnote{The Weyl transformation from $AdS_2\times
S^2$ to $\R^4$ introduces a boundary term for the action in $\R^4$
proportional to $\tr(\phi^I\phi^I)$.} We emphasize that we have
presented the analysis on $AdS_2\times S^2$ purely as a matter of
convenience. We also note that our result for the expectation value
applies to a general 't Hooft loop in any gauge theory where the
matter fields are not excited by the operator or where adjoint
matter fields have scale invariant singularities. We now proceed to
the study of the quantum definition of the operator.

\subsection{Quantum 't Hooft Loop}

The 't Hooft loop operator $T(\LR)$ is defined by integrating in the
path integral over all fields which have a prescribed singularity
near the loop.  In order to   evaluate the expectation value of the
't Hooft operator in the quantum theory, we must explicitly specify
the measure of integration in the path integral.

The path integral for the 't Hooft operator $T(\LR)$  is performed
by expanding the fields around the singularity
\ba
A&=A_0+\hat{A}\cr \phi^I&=\phi_0^I+\hat{\phi}^I\,,
\nonumber
\ea
where $(A_0,\phi_0^I)$ is the classical singularity corresponding to
a 't Hooft loop $T(\LR)$ (\ref{def-thooft}) and
$(\hat{A},\hat{\phi}^I)$ are the non-singular quantum fluctuations
that we must integrate over in the path integral.
In order to define the path integral and eliminate the gauge
redundancies, we must specify a gauge fixing procedure. We quantize
the theory in the background field gauge, where $(A_0,\phi_0^I)$ is
the background about which the path integral is expanded. The gauge
fixing condition we consider is the dimensional reduction to four
dimensions of the covariant background field gauge fixing condition
in ten-dimensional super Yang-Mills. It is given by
\ba
D_{0}^M\hat{A}_M=0\,,
\nonumber
\ea
where
\ba D_{0}^M=\partial^M -i[A_0^M,~\cdot~]\,. \nonumber \ea
In terms of the fields in ${\cal N}=4$ super Yang-Mills multiplet
the gauge fixing condition takes the form
\ba
D_{0}^\mu\hat{A}_\mu-i[\phi^I_0,\hat{\phi}^I\hskip+1pt ]=0\,.
\nonumber
\ea
The gauge fixing procedure requires introducing Faddeev-Popov ghosts
in the path integral as well as the addition of the following gauge
fixing term and ghost action to the ${\cal N}=4$ super Yang-Mills
action
\ba
S_{\rm gf}=\oo{g^2}\int d^4 x \sqrt h\, \tr \left[
  D^M_{0} \hat{A}_M
D^N_{0} \hat{A}_N  -  \bar c D^M_{0} D_M c
\right]\,,
\label{gfaction}
\ea
which in terms of four-dimensional fields reads
\ba
S_{\rm gf}=\oo{g^2}\int d^4 x \sqrt h\, \tr \left[
  \left(D^\mu_{0} \hat{A}_\mu-i[\phi^I_0,\hat{\phi}^I\hskip+1pt ]\right)^2
  -  \bar c D^\mu_{0} D_\mu c + \bar c [\phi^I_0,[\hat{\phi}^I,c]]
\right]\,.
 \label{gfactiona}
\nonumber
\ea
From the gauge fixed path integral  and by expanding around the
background created by the 't Hooft operator $T(\LR)$,  Feynman rules
can be extracted and the expectation value of $T(\LR)$ can be
computed to any desired order in perturbation theory. It is given by
the sum over all connected vacuum diagrams.

The definition given thus far for the 't Hooft operator  $T(\LR)$
is, however, not gauge invariant. The singularity produced by the
operator  $T(\LR)$
\ba
F={B\over 2}\, \hbox{vol}(S^2)+i g^2\theta {B\over
16\pi^2}{dt\wedge dr\over r^2}\, \qquad \phi=\f{B}{2r}{g^2\over
4\pi} |\tau|\,,
\nonumber
\label{def-thoofta}
\ea
breaks the $G$-invariance  of the ${\cal N}=4$ super Yang-Mills
action to invariance under a stability subgroup $H\subset G$. The
choice of $B\in \tfrak$, which characterizes the strength of the
singularity, determines the unbroken gauge group $H$. This is
generated by the generators $T\subset \gfrak$ for which
\ba
[B,T]=0\,.
\ea
In order to have a path integral definition of the 't Hooft operator
$T(\LR)$ which is gauge invariant, we must integrate over all the
$G$-orbits of $B\in \tfrak$ along the loop. This integration, which
we include in our definition of the path integral, restores
$G$-invariance. The integral we must perform is over the adjoint
orbit of $B$
\ba
 O(B)\equiv \{B_{\mathsf g}=\mathsf{g} B\mathsf{g}^\mo, \ \mathsf{g} \in G\}\,,
\ea
which is diffeomorphic to the coset space $G/H$. The integration
over the adjoint orbit of $B$ is reminiscent of the integration over
collective coordinates around a soliton in quantum field theory. In
the context of quantization of the 't Hooft operator, integration
over $O(B)$ follows from demanding that the path integral is gauge
invariant. In the computation of a general 't Hooft operator in an
arbitrary gauge theory we must also include this measure factor.

Having an explicit definition of the quantum 't Hooft operator
$T(\LR)$ we now proceed to calculate the expectation value of
$T(\LR)$ to one loop order. Integrating out the quantum fluctuations
to one loop requires expanding the complete gauge fixed ${\cal N}=4$
super Yang-Mills action obtained by combining  (\ref{10Daction}) and
(\ref{gfaction})    to quadratic order in the fluctuations. The
quadratic action   is given by the dimensional reduction to four
dimensions of
\ba
S_{(2)}&=& \oo{ g^2}\int d^4x \sqrt{h}\, \tr
\left[
    \hat{A}_M  (-\delta^{MN}D_0^2+2i F_{0}^{MN}
)\hat{A}_N
+   i\bar \psi  \Gamma^M D_{0M}  \psi -  \bar c D_0^2 c
\right]\,,\label{quad-action}
\nonumber
\ea
where we are packaging the ${\cal N}=4$ super Yang-Mills  fields
into ten-dimensional fields. Therefore, up to one loop order the
expectation value of the circular 't Hooft loop operator $T(\LR)$ is
given by\footnote{In Lorentzian signature the fermions are
Majorana-Weyl. In Euclidean signature, the fermions are chiral and
complex, but $\psi$ and $\bar{\psi}$ are not independent, resulting
in the exponent of $1/4$ for the fermionic
determinant.\label{Majorana}}
\ba
\langle T(\LR)\rangle_{G,\tau} = \exp\left({\tr(B^2) \over 8} g^2|\tau|^2\right)
\cdot{\left[\det_f\left(i\Gamma^M D_{0 M}\right)\right]^{1/4}\det_g\left(-D^2_{0}\right)\over \left[\det_b\left(-\delta^{MN}D_0^2+2i F_{0}^{MN}\right)\right]^{1/2}}\cdot \int d\mu_{O(B)}\,.
\label{oneloop}
\ea
The first factor arises, as we have seen in (\ref{circleresult}),
from the renormalized on-shell action evaluated on the classical
singularity produced by $T(\LR)$, the second one from integrating
out the fluctuations of the bosons, fermions   and ghost fields, and $\int d\mu_{O(B)}$ is
the integration over the adjoint orbit of $B$ required by gauge
invariance.

In Appendix \ref{cancellation} we show that the one loop
determinants all cancel among themselves. The reason behind this
cancellation is that the background for an 't Hooft loop operator
(\ref{theta-background}) is invariant under half of the
supersymmetries of the theory. Moreover, the background is self-dual
if we package the three components of the gauge field and the scalar
field $\phi$ sourced by the loop as a four component gauge
field.\footnote{This is the familiar statement that the monopole
equations arise by dimensional reduction to one lower dimension from
the self-duality equations, where   the scalar field in the monopole
equations arises from the fourth component of the gauge field.} The
cancellation of the determinants is then quite analogous to the
cancellation of the corresponding determinants of  ${\cal N}=4$
super Yang-Mills around an instanton background.

We now have to construct the metric on the adjoint orbit of $B$,
$O(B)$.\footnote{While $O(B)$ is diffeomorphic to $G/H$, their
metrics as a submanifold and as a quotient are different.} This is
obtained by computing
\ba
\tr (dB_{\mathsf g}^2)\,,
\nonumber
\ea
where $\mathsf{g}\in G$. This yields
\ba
\tr (dB_{\mathsf g}^2)=\tr\left([B,\mathsf{g}^{-1} d\mathsf{g}]^2\right)\,.
\label{adjoint}
\ea
We write the Lie algebra $\gfrak$ in the Cartan basis $\{{H_i}, {E_{\alpha}}\}$. The generators ${H_i}$ span the Cartan subalgebra $\tfrak\subset \gfrak$ and ${E_{\alpha}}$ are ladder operators associated to  roots $\alpha$ of the Lie algebra $\gfrak$. We can decompose the Maurer-Cartan form of the group $G$ in terms of the generators of $\gfrak$
 \ba
 \mathsf{g}^{-1} d\mathsf{g}=i\left(\sum_i d\xi^i {H_i}+\sum_{\alpha} d\xi^{\alpha} {E_{\alpha}}
\right)\,.
\label{maurer}
\ea
In order to explicitly determine the physical metric in $O(B)$ we must specify the overall normalization. We fix the normalization of the metric from the quadratic form defined by  the on-shell action (\ref{circleresult}) of  the 't Hooft operator.
  Therefore, by evaluating (\ref{adjoint}) the physical metric on the  adjoint orbit of $B$ is given by
    \ba
  ds^2_{O(B)}={g^2|\tau|^2\over 4}
\mathop{\sum_{\alpha>0}}_{\alpha(B)\neq 0} \alpha(B)^2\, 2\,\tr\left(E_\alpha E_{-\alpha}\right)|d\xi^\alpha|^2\,,
\label{orbit-metric}
\nonumber
\ea
where the sum is over all the positive roots $\alpha$ that do not
annihilate $B$, and we have used that
$[X,E_\alpha]=\alpha(X)E_\alpha$ for any $ X \in \tfrak$. This
implies that\footnote{As usual in path integrals, a factor of
$1/\sqrt{2\pi}$ multiplies each integration variable $d\xi^\alpha$,
which guarantees that the path integral for the Gaussian model is
normalized to $1$.}
 \ba
 \int d\mu_{O(B)}= \left({g^2|\tau|^2\over 8\pi}\right)^{{\dim(G/H)/ 2}}{\rm Vol}(G/H)
\mathop{\prod_{\alpha>0}}_{\alpha(B)\neq 0} \alpha(B)^2\,,
\label{orbit-volume}
 \ea
since
\ba \mathop{\sum_{\alpha>0}}_{\alpha(B)\neq 0}
2\,\tr\left(E_\alpha E_{-\alpha}\right)|d\xi^\alpha|^2=ds^2_{G/H}\,.
\nonumber \ea

 The complete one loop result for the expectation value of a circular 't Hooft operator $T(\LR)$ in ${\cal N}=4$ super Yang-Mills with arbitrary gauge group $G$ is then
 \ba
 \langle T(\LR)\rangle_{G,\tau} = \exp\left({\tr(B^2) \over 8} g^2|\tau|^2\right) \left({g^2|\tau|^2\over 8\pi}\right)^{{\dim(G/H)/ 2}}{\rm Vol}(G/H)
\mathop{\prod_{\alpha>0}}_{\alpha(B)\neq 0} \alpha(B)^2\,,
 \label{hooftfinal}
 \ea
  where we recall that $B$ is identified with the highest weight   $\LLw$ of the representation $\LR$ of $\LG$, which labels the operator.

Since we have given a complete definition of the path integral
measure  and have an explicit gauge fixed action, the expectation
value of the 't Hooft  operator $T(\LR)$ can now be computed to any
desired higher order  in perturbation theory. It is given by the sum
over all connected vacuum graphs around the singularity created by
$T(\LR)$.

  \medskip
  \noindent
  {\it Examples}

The discussion thus far has been very general, applying to an
arbitrary  circular 't Hooft operator   $T(\LR)$ in ${\cal N}=4$
super Yang-Mills with gauge group $G$. Such an operator is labeled
by a representation $\LR$ of $\LG$. In order to make the discussion
a bit less abstract, here we present the relevant formulas for
various elementary gauge groups.

 \noindent $\bullet$ $G=SU(2)$ and $SO(3)$.  't Hooft operators in this theory  are labeled by a highest weight of the dual group, which are $\LG=SO(3)$ and $\LG=SU(2)$ respectively. A highest weight of $SO(3)$ can be labeled in terms of  a  spin  $j\in \Z^+$ while for $SU(2)$ $j\in (1/2) \Z^+$. For an 't Hooft  operator with $j\neq 0$, the broken symmetry near the loop is $H=U(1)$  (for $j=0$, we just get the identity operator). In this case
  \ba
 \langle T(j)\rangle_{G,\tau} =
\exp\left(\f{j^2}{ 4} g^2|\tau|^2\right)
 j^2g^2|\tau|^2
 \,.  \label{thooft-j}
\ea

  \noindent $\bullet$ $G=U(N)$.  't Hooft operators in this theory  are labeled by a highest weight of the dual group, which is also $\LG=U(N)$. A highest weight of $U(N)$ can be labeled by a set of integers $\LLw=[m_1,m_2,\ldots,m_N]$ with $m_1\geq m_2\geq\ldots \geq m_N$.\footnote{A highest weight of $U(N)$ with $m_N\geq 0$ is in one-to-one correspondence with a Young diagram containing $m_l$ boxes in the $l$-th row.}
The corresponding data characterizing the monopole singularity (\ref{def-thooft})  is given by
  \ba
B=\left(\begin{array} {cccc} m_1& 0&  \dots & 0 \cr
                                  0  &m_2&  \dots & 0 \cr
                \vdots  & \vdots & \ddots& \vdots\cr
            0&0  &\dots & m_N\cr
            \end{array}
            \right)\in \tfrak\simeq \ufrak(1)^N\,.
            \nonumber
  \ea
Let us now consider various representations of $U(N)$:

    $\vartriangleright \LLw=[k,0,\ldots, 0]$. This corresponds to the rank-$k$ symmetric representation. The stability group in this case is $H=U(1)\times U(N-1)$ and
    \ba
 \langle T\left([k,0,\ldots, 0]\right)\rangle_{G,\tau}
=
\exp\left({k^2 \over 8} g^2|\tau|^2\right) \left({g^2|\tau|^2
k^2\over 4}\right)^{N-1}  {1\over (N-1)!} \,,
 \ea
where we have used that ${\rm Vol}(U(N))=
(2\pi)^{N(N+1)/2}/\prod_{n=1}^{N-1}n!$ \cite{1980InMat..56...93M}.

    $\vartriangleright \LLw=
[\stackrel{k\, {\rm times}}{\overbrace{1,\ldots,1}}
,0,\ldots,0]
$. This corresponds to the rank-$k$ antisymmetric representation. The stability group in this case is $H=U(k)\times U(N-k)$ and
    \ba
 \langle T\left([1,\ldots,1,0,\ldots,0]\right)\rangle_{G,\tau}= \exp\left({k \over 8} g^2|\tau|^2\right) \left({g^2|\tau|^2\over 4}\right)^{k(N-k)}  {\prod_{n=1}^{k-1} n!\over \prod_{n=1}^{k} (N-n)!}\,.
 \ea
\vfill\eject
   $\vartriangleright \LLw=[m_1,m_2,\ldots,m_N]$ with $m_1>m_2\ldots > m_N$.   The stability group in this case is $H=U(1)^N$ and
    \ba
&& \langle T
\left([m_1,m_2,\ldots,m_N]\right) \rangle_{G,\tau}
\nn\\
&& \hskip 30pt= \exp\left({\sum_i m_i^2 \over 8} g^2|\tau|^2\right) \left({g^2|\tau|^2\over 4}\right)^{N(N-1)/2}  {1\over \prod_{n=1}^{N-1} n!}\prod_{i<j}(m_i-m_j)^2.
\label{thooft-gen}
 \ea


\section{Wilson loop expectation value}
\label{sec-wilson}

The aim of this section is to compute the expectation value of the
circular Wilson loop in ${\cal N}=4$ super Yang-Mills at strong
coupling. The ultimate goal  is to show that our  result
(\ref{hooftfinal}) for the expectation value of the circular  't
Hooft operator  at weak coupling maps  in the dual theory to the
expectation value of the Wilson loop at strong coupling, thereby
exhibiting $S$-duality in ${\cal N}=4$ super Yang-Mills for
correlation functions.

The supersymmetric circular Wilson loop in ${\cal N}=4$ super
Yang-Mills with gauge group $G$ is labeled by a representation $R$
of $G$. It is given by \cite{Rey:1998ik,Maldacena:1998im}
\ba W(R)\equiv \Tr_R {\rm P} \exp{{\oint(iA+\phi)}}\,, \nonumber
\ea
where $\phi\equiv \phi^I n^I$ and $(n^I)$ is a unit vector in
$\R^6$.

A remarkable property of the supersymmetric circular Wilson loop
$W(R)$ is that its expectation value can be computed in terms of a
matrix model, thereby reducing the complexity of the path integral
of a four-dimensional field theory to a matrix integral. This result
was first conjectured in \cite{Erickson:2000af,Drukker:2000rr}, and
was based on  computations of the Wilson loop in perturbation
theory. This remarkable result has   been proven in an elegant paper
by Pestun \cite{Pestun:2007rz}, who, using localization techniques,
has shown that the path integral over the four-dimensional fields
reduces to an integral over a zero mode, which  corresponds to the
variable of integration in the matrix model integral.

The expectation value of the supersymmetric circular Wilson loop $W(R)$ transforming in a representation $R$ of $G$ is given by the matrix integral \cite{Erickson:2000af,Drukker:2000rr,Pestun:2007rz}
\ba
\langle W(R)\rangle_{G,\tau}=\oo{{\cal Z}}\int_{ \gfrak} [dM] \exp\left({-\f{2}{g^2} \tr(M^2)}\right)
\Tr_R \,e^{M}\,.
\label{M-integral}
\ea
$M$ is an element in the Lie algebra $\gfrak$ corresponding to $G$ and ${\cal Z}$ is the matrix model model partition function. As in the previous section, $\tr(\ ,\ )$ is the invariant metric on the Lie algebra $\gfrak$, and is normalized so that the length-squared of the short coroots is two. This normalization fixes the measure $[dM]$, which is the volume element on the Lie algebra $\gfrak$. The normalization factor is
\ba
{\cal Z}=\int_\gfrak [dM] \exp\left(-\f{2}{g^2}\tr(M^2)\right)
=\left({\pi g^2\over 2}\right)^{\dim (G)/2}\,.
\label{partition-function}
\ea

We now ``gauge fix" and reduce the integral over $\gfrak$ to
integration over the Cartan subalgebra $\tfrak$. Any $M\in \gfrak$
is conjugate to an element $X$ in the maximal torus $T$ of $G$, and
the Lie algebra $\gfrak$ decomposes into orbits of the $G$-action,
with the generic orbit being diffeomorphic  to $G/T$. In formulas
\ba \forall M\in \gfrak\, ,\ \exists \,X\in T, \ \mathsf{g}\in G/T :
\qquad\qquad  M= \mathsf{g}X\mathsf{g}^{-1}\,, \nonumber \ea
and the metric in $\gfrak$ is given by
\ba
\tr(dM^2)=\tr(dX^2)+\tr[X,\mathsf{g}^{-1} d\mathsf{g}]^2\,.
\nonumber
\ea
Using the decomposition of the Maurer-Cartan form of $G$ in
(\ref{maurer}) we find that
\ba
\tr(dM^2)= \tr(dX^2)+
\mathop{\sum_{\alpha>0}}_{\alpha(X)\neq 0}
\alpha(X)^2\, 2\,\tr\left(E_\alpha
E_{-\alpha}\right)|d\xi^\alpha|^2\,,
\nonumber
\ea
where $\alpha$ are the roots of $\gfrak$.

Generically, there is more than one $X$ in the maximal torus $T$
associated with a given $M\in \gfrak$, but these are related to each
other by the action of the Weyl group $\Wcal$ of $\gfrak$.
Correspondingly, the orbits of the $G$-action are parametrized by
$X\in T$ up to the action of $\Wcal$.  Therefore, integration over
the orbit yields
\ba \int_\gfrak [dM] e^{-\f{2}{g^2} \tr(M^2)}\Tr_R e^M =\f{{\rm
Vol}(G/T)}{|\Wcal|} \int_{\mathfrak t} [dX] \Delta(X)^2
 e^{-\f{2}{g^2}\langle X, X\rangle}
\Tr_R e^X
\,,
\label{reduced}
\ea
where
\ba
\Delta(X)^2= \prod_{\alpha}|\alpha(X)|
=\prod_{\alpha>0} \alpha(X)^2\,,
\nonumber
\ea
and  $\langle\  ,\ \rangle$ is the metric on the Cartan subalgebra
$\tfrak$. The factor of  $\Delta(X)^2$ plays the role of the
Vandermonde determinant in Hermitian matrix models.

It is convenient to write the insertion of the group character  in
(\ref{reduced}) as the sum over all the weights $v\in \Omega(R)$ in
the representation $R$
\ba \Tr_R \,e^X=\sum_{v\in \Omega(R)} n(v)\, e^{v(X)}, \nonumber \ea
where $n(v)$ is the multiplicity of the weight $v$   and $\Omega(R)$
is the set of all weights in the representation $R$. By completing
squares in the exponential, we obtain
\ba
\langle W(R)\rangle_{G,\tau}\hskip-3pt
&=&
\f{{\rm Vol}(G/T)}{|\Wcal|{\cal Z} }
\sum_{v\in \Omega(R)} n(v)\, e^{\f{g^2 }8\langle v, v \rangle}
\nn\\
&&~~~~~~\times\int_\tfrak [dX]
 e^{-\f{2}{g^2}\langle X, X\rangle}
\prod_{\alpha>0}\left(\alpha(X)+\f{g^2}4 \langle \alpha, v\rangle\right)^2.
\label{W0}
\ea
For each weight $v\in  \Omega(R)$, we obtain the expectation value
of a polynomial in the $X$'s, which can be evaluated using Wick
contractions, yielding a polynomial in the coupling constant $g$.

Since we are interested in understanding the action of $S$-duality
on our perturbative computation of the 't Hooft operator
(\ref{hooftfinal}), we need to solve the matrix model for the Wilson
loop at strong coupling. The large $g$ behaviour of the Wilson loop
(\ref{W0}) is controlled by the exponential prefactor. At large $g$,
the leading  contribution arises from the terms in the sum over
weights involving the longest weights in the representation $R$. It
is for these weights that the length of the weight -- given by
$\langle v,v \rangle$ -- is maximal.

The longest weights $v\in  \Omega(R)$ are related to the highest
weight\footnote{The highest weight  appears with multiplicity one in
the set $\Omega(R)$ of all possible weights in the representation,
as otherwise the representation would be reducible.} in the
representation $R$ - which we denote by $w$ -- by the action of the
Weyl group $\Wcal$. However,  there is an invariant subgroup
$H\subset G$ that leaves that  highest weight $w$ invariant, and the
Weyl group of $H$ -- which we denote by $\Wcal(H)$ --   acts
trivially on $w$.

The strong coupling limit of the circular Wilson loop operator is
thus given by
\ba \langle W(R)\rangle_{G,\tau} &=& \f{{\rm Vol}(G/T)}{
|\Wcal(H)|{\cal Z}} e^{\f{g^2}8 \langle w,w\rangle} \int_{\tfrak}
[dX]
 e^{-\f{2}{g^2}\langle X, X\rangle}
\prod_{\alpha>0}\left(\alpha(X)+\f{g^2}4 \langle \alpha, w\rangle\right)^2\,.
\label{W1}
\nonumber
\ea
The leading contribution at strong coupling is obtained  by
factoring  out $\langle \alpha, w\rangle$ from the integral for the
roots $\alpha$ in the Lie algebra $\gfrak$ for which $\langle
\alpha, w\rangle\neq 0$. This yields
\ba \langle W(R)\rangle_{G,\tau} &=&\f{{\rm Vol}(G/T)}{
|\Wcal(H)|{\cal Z}} e^{\f{g^2}8 \langle w,w\rangle} \Bigg
(
\prod_{\alpha>0, \langle \alpha, w\rangle\neq 0}
\f{g^2}4 \langle \alpha, w\rangle
\Bigg
)^2
\nn\\
&&~~~~~~~~~~~~~~~~~~~~~
\times
\int_{\tfrak} [dX]
 e^{-\f{2}{g^2}\langle X, X\rangle}
\mathop{\prod_{\alpha>0}}_{\langle \alpha, w\rangle= 0}
\alpha(X)^2
\,.\label{W2}
\ea
We can now perform the integral over the Cartan subalgebra elements
$X$,  which is proportional to the inverse of ${\rm Vol}(H/T)$ (see
Appendix \ref{volume} for details). The expectation value of the
circular Wilson loop $W(R)$ in ${\cal N}=4$ super Yang-Mills at
strong coupling is then given by
\ba
\langle W(R)\rangle_{G,\tau}
= \exp\left({\langle w,w\rangle \over 8} g^2 \right) \left({g^2 \over 8\pi}\right)^{{\dim(G/H)/ 2}}{\rm Vol}(G/H)\prod_{\alpha>0, \langle \alpha,w\rangle \neq 0} \langle \alpha, w\rangle^2\,,
 \label{leading-mixed2-1}
\ea
where $w$ is the highest weight of the representation $R$ of
$G$.\footnote{We evaluate (\ref{leading-mixed2-1}) for some sample
representations and gauge groups  in Appendix \ref{ex-wil}.}


\section{\texorpdfstring{$S$}{S}-duality for loop operators}
 \label{duality}

${\cal N}=4$ super Yang-Mills with gauge group $G$ is conjectured to
have a  symmetry group $\Gamma\subset SL(2,\R)$, which acts  on all
the gauge invariant operators in the theory as well as on the
complexified coupling constant
 \ba
 \tau={\theta\over 2\pi}+{4\pi i\over g^2}\,,
 \nonumber
 \ea
on which it acts by fractional linear transformations
\ba
\tau\rightarrow  {a\tau+b\over c\tau+d}\,,\qquad \qquad\qquad
\left(\begin{array} {cc} a& b\cr c  &d\cr
\end{array}\right)
           \in SL(2,\R)\,.
           \nonumber
\ea
The symmetry group $\Gamma$ has two generators, usually denoted by  $T$ and $S$. $T$ generates the
classical symmetry
\ba
T: \tau\rightarrow \tau+1\,,
\nonumber
\ea
which follows by inspecting   the ${\cal N}=4$ super Yang-Mills  path integral.
  $S$ conjecturally generates a    quantum symmetry which exchanges the gauge group $G$ with the dual group $\LG$ and inverts the coupling constant
\ba
S:\tau\rightarrow -{1\over n_\gfrak \tau}\,,
\label{actioncoupl}
\ea
where $n_\gfrak$ is the ratio $|\hbox{long root}|^2/|\hbox{short root}|^2$
for $\gfrak$ (see footnote \ref{ng}). When the Lie algebra is simply laced $\Gamma=SL(2,\Z)$.

In ${\cal N}=4$ super Yang-Mills with gauge group $G$, a Wilson
operator is labeled by a representation $R$ of $G$ while an 't Hooft
operator is labeled by a representation $\LR$ of the dual group
$\LG$. Under the action of $S$-duality a Wilson  operator  in the
theory with gauge group $G$   maps to an 't Hooft operator in the
theory with the dual group $\LG$ and vice versa
  \ba
  \begin{tabular}{ccc}
  $G$& &$\LG$\cr
 $W(R)$&$\longleftrightarrow$ & $T(R)$\cr
 $T(\LR)$&$\longleftrightarrow$ & $W(\LR)$
  \end{tabular}
  \nonumber
  \ea
Non-trivial evidence for $S$-duality was presented by
\cite{Kapustin:2005py}, where it was  shown that   given a Wilson
operator in the theory with gauge group $G$ that   one can construct
the  classical singularity of an 't Hooft operator for the theory
with dual gauge group $\LG$ with precisely the same quantum numbers
as the original Wilson operator. The $S$-duality conjecture  further
predicts that the correlation functions of dual operators are the
same. In particular, $S$-duality predicts that the expectation value
of an 't Hooft operator gets mapped to the expectation value of a
Wilson operator in the dual theory
\ba \langle T(\LR)\rangle_{G,\tau}= \langle
W(\LR)\rangle_{\LG,\Ltau}\,. \label{predict}
\ea

We now use  our computation of the semiclassical 't Hooft operator
expectation value, and of the expectation value of the Wilson
operator strong coupling to exhibit   that correlation functions in
${\cal N}=4$ super Yang-Mills transform precisely as predicted by
$S$-duality. We recall that up to one loop order, the expectation
value of a circular 't Hooft loop operator in ${\cal N}=4$ super
Yang-Mills with gauge group $G$ is given by (\ref{hooftfinal})
\ba
\langle T(\LR)\rangle_{G,\tau} = \exp\left({\tr(B^2) \over 8}
g^2|\tau|^2\right) \left({g^2|\tau|^2\over 8\pi}\right)^{{\dim(G/H)/
2}} \hspace{-2mm} {\rm Vol}(G/H) \hspace{-2mm}
\mathop{\prod_{\alpha>0}}_{\alpha(B)\neq 0} \alpha(B)^2\,,
\label{hooftfinala}
\ea
where $B$ is the highest weight of the representation $\LR$ of
$\LG$. The dual operator is a circular Wilson loop in ${\cal N}=4$
super Yang-Mills with gauge group $\LG$, whose expectation value at
strong coupling is given by\footnote{Note that in Section $3$ we
calculated the Wilson loop for gauge group $G$, while here we need
the result for gauge group $\LG$. This explains the appearance of
the dual group, dual coupling and so on.}
\ba
&&\langle W(\LR)\rangle_{\LG,\Ltau}
\nn\\
&&
\hspace{5mm}
= \exp\left({\langle \LLw,\LLw\rangle \over 8} \LLgc^2 \right) \left({\LLgc^2 \over 8\pi}\right)^{{\dim(\LG/\LH)/ 2}}
{\rm Vol}(\LG/\LH)
\mathop{\prod_{\Lalpha >0}}_{\langle \Lalpha,\Lw\rangle \neq 0} \langle\Lalpha, \LLw\rangle^2\,,
 \label{leading-mixed2-2}
\ea
where $\LLw$  is the highest weight of the representation $\LR$ of $\LG$.

In order to study the prediction of $S$-duality for these
correlators, we first note that the action of $S$-duality on the
coupling constant (\ref{actioncoupl}) implies that
 \ba
 (\LLgc)^2=n_\gfrak\, g^2|\tau|^2\,.
 \label{map1}
 \ea
We should also  pay attention to the difference between $\tr(B^2)$
and $\langle \LLw,\LLw\rangle$ when comparing the  correlators,
since they are constructed in terms of the  metric defined on
$\tfrak$ and ${}^L\tfrak^*$  respectively. These metrics are in turn
induced from the metrics on $\gfrak$ and $\Lgfrak$. In our
computations, we have normalized the Lie algebra metrics that appear
in the Lagrangians so that short coroots have length-squared equal
to two, or equivalently long roots have length-squared equal to two.
However,  roots in $G$ are identified with coroots of  the dual
group $\LG$, so a long root in $G$ is identified with a long coroot
in $\LG$. It follows that when we identify $\tfrak$ with
${}^L\tfrak^*$, the metric on $\tfrak$ is $n_\gfrak$ times the
metric on ${}^L\tfrak^*$. Therefore,  the norms of $B$ and $\LLw$
are related by\footnote{\label{metrics} More precisely, the
isomorphism $\varphi: \LLt^* \ra \mathfrak t$ satisfies
$B=\varphi({\LLw})$, $\langle \varphi({\LLw}),\varphi({\LLw})\rangle
=n_\gfrak \langle {\LLw},{\LLw}\rangle.$ }
\ba
 \tr(B^2)=n_\gfrak\langle\LLw,\LLw\rangle\,.
 \label{normrelate}
\ea
Finally, we can relate ${\rm Vol}(G/H)$ and ${\rm Vol}(\LG/\LH)$
(see Appendix \ref{volume}):
\ba {\rm Vol}(G/H) \mathop{\prod_{\alpha>0}}_{ \alpha(B)\neq 0}
\alpha(B)^2=n_\gfrak^{\dim(G/ H)/2}\, {\rm Vol}(\LG/\LH)
\mathop{\prod_{{}^L\negthinspace\alpha>0}}_{
\langle{}^L\negthinspace\alpha,\Lw\rangle \neq 0} \langle \LLalpha,
\LLw\rangle^2\,.
 \label{map2}
\ea

Inserting equations (\ref{map1}, \ref{normrelate}, \ref{map2}) into
the formula for the expectation value of the Wilson loop at strong
coupling (\ref{leading-mixed2-2}), we find   precisely the same
result we obtained for the expectation value of the  't Hooft
operator (\ref{hooftfinala}) in the dual theory.

To summarize, we have shown up to the order we have computed that
the expectation value of a Wilson operator is exchanged under
$S$-duality  with the expectation value of an 't Hooft operator,
thus exhibiting that these correlation functions in ${\cal N}=4$
super Yang-Mills transform precisely as predicted by the $S$-duality
conjecture.


\section{Saddle points from monopole screening}
\label{bubbling}

In the computation of the expectation value of the circular Wilson
loop operator $W(\LR)$ in ${\cal N}=4$ super Yang-Mills with gauge
group $\LG$, we have represented\footnote{Note that we are
considering a Wilson loop in the theory with gauge group $\LG$,
which explains the appearance of the dual representation, dual
coupling and so on when compared to Section $3$.} the insertion of
the  group character in the matrix integral in terms of the sum over
all   weights $\LLv\in\Omega(\LR)$ in the representation $\LR$ of
$\LG$
\ba \Tr_{\LR} \, e^X=\sum_{{}^L\negthinspace v\in \Omega(\LR)}
n(\hspace{-.5mm} \raisebox{.8mm}{\mbox{\it \scriptsize{L}}}v
\hspace{-.2mm}) \, e^{{}^L\mspace{-2mu} v(X)},
\nonumber
\ea
where $ n(\hspace{-.5mm} \raisebox{.8mm}{\mbox{\it \scriptsize{L}}}v
\hspace{-.2mm}) $ is the multiplicity of the weight $
\raisebox{.8mm}{\mbox{\it \scriptsize{L}}}v $   and $\Omega(\LR)$ is
the set of all weights in the representation $\LR$.

We have noted that the leading contribution at strong coupling
$\LLgc \gg 1$ arises from the longest weights, those with maximal
norm $\langle \LLv,\LLv\rangle$. These in turn   are obtained from
the highest weight  of the representation  $\LLw\in \Omega(\LR)$ by
the action of the Weyl group.

It is instructive to  also consider  the effect  of the non-longest
weights in the representation $\LLv\in\Omega(\LR)$ to the
expectation value of the Wilson loop. The  leading contribution  of
a non-longest weight $\LLv\in\Omega(\LR)$ at strong coupling  is
proportional to
\ba
\exp\left({\langle \LLv,\LLv\rangle \over 8} \LLgc^2 \right)
\left({\LLgc^2 \over 8\pi}\right)^{\half\dim(\LG/\LH(\LL v))}{\rm
Vol}(\LG/\LH(\LLv)) \mathop{\prod_{{}^L\negthinspace \alpha>0}}_{
\langle{}^L\negthinspace\alpha,\Lv\rangle \neq 0} \langle \LLalpha,
\LLv\rangle^2\,,
 \label{leading-mixed3}
\ea
where $\LH(\mspace{-2mu}\LLv)$ is the subgroup of $\LG$ that leaves
the weight $\LLv$ invariant. In the strong coupling limit,   this
contribution   is exponentially suppressed with respect to the
contribution   from the longest weights, which is proportional to
$\exp[\langle \LLw, \LLw\rangle\LLgc^2/8] \gg \exp[\langle \LLv,
\LLv\rangle\LLgc^2/8]$.

$S$-duality predicts that these subleading contributions to the
Wilson loop at strong coupling should also arise in the
semiclassical computation of the expectation value of the 't Hooft
operator. We argue that these subleading exponentials appear in the
't Hooft loop correlator  via the physics of monopole screening.

The physics of screening of an 't Hooft operator by a regular
monopole is $S$-dual to the more familiar screening of an electric
source by dynamical gluons.  The non-abelian charge inserted by a
Wilson loop in a representation $\LR$ of $\LG$ can be screened by
gluons, which also carry non-abelian charge  and are constantly
appearing and disappearing from the vacuum due to quantum
fluctuations. The gluons are, however, uncharged under the center
$Z(\LG)\subset \LG$. Therefore, the quantum number   that cannot be
screened is the charge  of the representation $\LR$ of the Wilson
loop under $Z(\LG)$. On the other hand, since $\pi_1(G)\simeq
Z(\LG)$, the quantum number associated with the $S$-dual 't Hooft
operator $T(\LR)$ that cannot be screened by regular monopoles is an
element of $\pi_1(G)$, which is precisely the topological charge
carried by an 't Hooft operator \cite{'tHooft:1977hy}.

In the path integral definition of a t 'Hooft loop $T(\LR)$ we have
quantized the singularity produced by $T(\LR)$ in the background
field gauge. The singularity   determining the field configuration
near the loop is specified by the highest weight $\LLw$, which gets
identified with $B$. The norm of $B$ -- $\tr(B^2)$ -- determines the
strength of the singularity. The subleading exponentials predicted
by $S$-duality arise from boundaries of the region of integration in
field space where the singularity produced by the 't Hooft operator
$T(\LR)$ is weaker, and is controlled by a non-longest weight $\LLv$
of the representation $\LR$.  The necessity to include the less
singular configurations was noticed in \cite{Kapustin:2006pk}, where
the phenomenon was dubbed ``monopole bubbling''. This weaker
singularity arises physically from the physics of monopole
screening, whereby a regular 't Hooft-Polyakov monopole approaches a
singular monopole (an 't Hooft operator), and screens the charge of
the 't Hooft operator.

The charges of regular 't Hooft-Polyakov monopoles are spanned by
the simple coroots of the Lie algebra $\gfrak$, which generate the
coroot lattice. When we bring a regular monopole -- labeled by a
coroot --  near an 't Hooft operator $T(\LR)$, the charge of the 't
Hooft operator is screened. The resulting effective charge is
obtained by the action of lowering operator associated with the
coroot labeling the regular monopole on the highest weight $\LLw$
that characterizes the singularity of the 't Hooft operator
$T(\LR)$. The action of the ladder operators associated with the
regular monopoles on the highest weight $\LLw$ generates all the
weights in the representation $\LR$ \cite{Kapustin:2006pk}.

This was explicitly realized in \cite{Cherkis:2007jm} by
constructing a classical solution to the equations of motion that contains a regular monopole in the
presence of a singular monopole in the cases $G=SU(2)$ and
$G=SO(3)$. When the gauge group is $G=SU(2)$, the dual group is
$\LG=SO(3)$, and the minimal 't Hooft operator $T(1)$ carries spin one
with respect to $\LG=SO(3)$.  In the spin one representation a
state with vanishing weight can be obtained from the highest weight
state by applying the lowering operator. This can be translated into
the language of monopole screening by  noting that $T(1)$ can be
completely screened by an 't Hooft-Polyakov monopole.  Indeed the
singularity in the solution disappears when the regular monopole
approaches the 't Hooft operator.  If on the other hand $G=SO(3)$,
the minimal 't Hooft operator $T(1/2)$ carries spin one-half with
respect to $\LG=SU(2)$.  The spin one-half representation has no
state with vanishing weight and so the 't Hooft operator cannot be
screened. In the solution, the size of the regular monopole remains
finite as it approaches the singularity, and the strength of the
singularity remains intact/unscreened.

The subleading saddle points corresponding to the non-longest
weights in the representation should then be included in the path
integral and can be computed in the same manner as we have done in
Section $2$. Instead of quantizing the singularity of the 't Hooft
operator $T(\LR)$ associated with the highest weight of $\LR$, we
quantize the singularity produced by the weaker singularities that
appear at the boundaries  of the region of integration of field
space, which are labeled by the weights of  $\LR$. These saddle
points are then in one-to-one correspondence with the subleading
contributions to the expectation value of the Wilson loop in
(\ref{leading-mixed3}), and reproduce the Wilson loop result
including the prefactor.


\section{Discussion}

We conclude by summarizing our results and describing several
interesting lines of inquiry stemming from this work. First we have
defined  the renormalized 't Hooft operator in terms of a path
integral quantized in the background field gauge around a certain
codimension three singularity created by the operator. We have shown
that an important ingredient that goes into  the definition of the
operator is the measure of integration in the path integral, which
is dictated by gauge invariance, and which requires integrating over
the adjoint orbit of the classical singularity produced by the 't
Hooft operator.  This measure factor    should be included in the
computation of a general 't Hooft operator in an arbitrary gauge
theory. We have then explicitly computed the expectation value of
the circular 't Hooft loop operator in ${\cal N}=4$ super Yang-Mills
up to one loop order. Computations to higher   orders in
perturbation theory can now be carried out by summing over all
connected vacuum diagrams generated by the path integral.

By solving for the expectation value of the   $S$-dual Wilson
operator at strong coupling we have been able to exhibit that
correlation functions in ${\cal N}=4$ super Yang-Mills transform
properly under $S$-duality.  We have shown that the perturbative
result for the expectation value of the 't Hooft operator up to one
loop order exactly reproduces the strong coupling expansion  of the
$S$-dual Wilson loop in the dual theory. Unlike  most of the
previous studies of $S$-duality, the matching goes beyond comparing
``topological'' features like the spectra, quantum numbers and so
on, but it rather tests the quantum dynamics underlying 
$S$-duality.\footnote{Note that our test of $S$-duality is purely
in field theory.  The tests that are based on AdS/CFT
and the identification of $S$-dualities in $\Ncal=4$ super-Yang Mills
and type IIB superstring theory
include \cite{Drukker:2000rr,Chen:2006iu,Gomis:2008qa}.
}

We have also argued that the subleading exponential corrections to
the expectation value of the Wilson loop at strong coupling can be
identified with the weaker singularities that appear near an 't
Hooft loop due to monopole screening. It would be interesting to
understand in more detail the contribution from these subleading
saddle points. In this respect, it would be illuminating to try to
evaluate the path integral for the circular 't Hooft loop using
localization techniques, extending to monopole operators the work by
Pestun \cite{Pestun:2007rz} for Wilson loops. We expect that
the subleading saddle points arise in this context as  solutions to
the localization equations. Furthermore, $S$-duality predicts that
the expectation value of the circular 't Hooft operator in ${\cal
N}=4$ super Yang-Mills is also described by a matrix model (see
equation (\ref{W0})), so it would be desirable to give a direct
derivation of this matrix model from the 't Hooft loop path
integral.

A worthwhile future direction is to extend our computation  for the
't Hooft loop expectation value to other gauge theories, in
particular to  finite ${\cal N}=2$ theories. An   interesting class
of ${\cal N}=2$ superconformal field theories are those that cannot
be obtained by quotienting ${\cal N}=4$ super Yang-Mills, such as
${\cal N}=2$ $SU(N)$ super Yang-Mills   coupled to $2N$ fundamental
hypermultiplets \cite{Seiberg:1994aj,Argyres:1994xh}, which are
conjectured to be invariant under an  $S$-duality group
$\Gamma\subset SL(2,\Z)$ \cite{Seiberg:1994aj,Argyres:1997cg} and to
exhibit rich duality relations at strong coupling
\cite{Argyres:2007cn}. ${\cal N}=2$  orbifolds \cite{Douglas:1996sw}
of ${\cal N}=4$ super Yang-Mills are also of  interest, and  Wilson
and 't Hooft operators in these theories are relevant probes for the
very rich   $S$-duality groups in these theories, where for
instance, the $S$-duality group of the $\hat{A}_{n-1}$ quiver gauge
theory is conjectured to be  the mapping class group of a torus with
$n$ punctures \cite{Witten:1997sc}. Very recently similar duality
relations were conjectured for a larger class of $\Ncal=2$ conformal
theories with more than one gauge group \cite{Gaiotto:2009we}.
Moreover gravity duals of such $\Ncal=2$ theories have been proposed
\cite{Gaiotto:2009gz}. Computing correlators of 't Hooft operators
and Wilson operators provide a useful framework to explore the
conjectured $S$-duality maps as well as  the holographic
correspondences.

It is also of interest to go beyond the computation  of the
expectation value,  and determine whether the perturbative
correlators of 't Hooft operators with local operators  get  mapped
to the corresponding strong coupling correlators of the $S$-dual
operators in the $S$-dual theory \cite{work}, by generalizing the
large $N$ results in \cite{Gomis:2008qa} to finite rank.  One can
also extend the computations in the present paper to correlators of
disorder operators in three-dimensional ${\cal N}=6$ superconformal
Chern-Simons theories \cite{Aharony:2008ug}, such as monopole
operators \cite{Borokhov:2002ib,Borokhov:2002cg} and vortex loop
operators \cite{Drukker:2008jm}. Another rich class of operators
that deserve further study are the mixed Wilson-'t Hooft loop
operators in ${\cal N}=4$ super Yang-Mills \cite{Kapustin:2005py}.
These operators insert dyonic probe particles and have interesting
conjectured transformation properties under $S$-duality
\cite{Kapustin:2005py}. Giving a quantum definition of these
operators and studying their correlation functions opens a novel
arena in which to probe the quantum dynamics underlying $S$-duality.

Wilson and 't Hooft operators exhibit the area law in the confining
and Higgs phases respectively, and are order parameters for these
phases. Is there a similar interpretation for the tree-level result
(\ref{circleresult}), which applies to the  't Hooft loop of any
gauge theory in the (Abelian or non-Abelian) Coulomb phase? A
notable feature is its dependence on the theta angle $\theta$.
Therefore it can be used to distinguish phases of a theory that have
different values of $\theta$. Precisely such phases for gauge
theories with $U(1)$ gauge group have been discussed recently in the
condensed matter literature. The orbital motion of electrons has
been shown to generate non-zero $\theta$ \cite{Hehl:2007jy}. The
so-called $\Z_2$ topological insulators in 3 + 1 dimensions are
particularly interesting examples, where time reversal symmetry sets
$\theta=\pi$ \cite{Qi:2008ew,Essin:2008rq}. Thus the expectation
value (\ref{circleresult}) distinguishes the topologically
non-trivial phase at $\theta=\pi$ from the vacuum at
$\theta=0$.\footnote{For a Wilson-'t Hooft operator $WT_{m,n}$
carrying electric and magnetic charges $(m,n)$, the expectation
value in the $\theta=0$ and $\theta=\pi$ phases are respectively \ba
\langle WT_{m,n} \rangle_{\theta=0}=\exp\left({g^2\over
8}m^2+{2\pi^2\over g^2}n^2\right), \qquad \hbox{and}\qquad \langle
WT_{m,n} \rangle_{\theta=\pi}=\exp\left({g^2\over 8}\left(m+{n\over
2}\right)^2+{2\pi^2\over g^2}n^2\right).\nonumber \ea}

Explicitly showing that the vacuum expectation value of 't Hooft
operators are exchanged under duality with the correlation function
of Wilson operators is a step in the right direction towards the
goal of finding the electric-magnetic duality transformation that
relates the two dual descriptions. In the past,  there have been
attempts to formulate Yang-Mills theories directly in terms of gauge
invariant variables,  i.e. Wilson variables. In this formulation of
the theory, the non-perturbative information of the theory is
encoded in the loop equation, which describes the dynamics of Wilson
loop operators in loop space. Constructing the loop equation for the
't Hooft loop variables and  studying how it maps to   the Wilson
loop equation, may provide a non-perturbative framework in which to
study the transformation between  gauge invariant electric and
magnetic variables that underlies $S$-duality in ${\cal N}=4$ super
Yang-Mills.

 
\medskip

\subsection*{Acknowledgments}

We would like to thank Evgeny Buchbinder, Nadav Drukker, Simone
Giombi, Anton Kapustin, Shunji Matsuura, Yusuke Nishida, Vasily
Pestun, Ashoke Sen, Shinsei Ryu, Kostas Skenderis, Arkady Vainshtein, Peter van
Nieuwenhuizen, Herman Verlinde and Edward Witten for helpful
comments. J.G. reported the results of this paper at the KITP
Conference ``Dualities in Physics and Mathematics'', and would like
to thank the KITP for its hospitality and for organizing a very
stimulating conference, which was supported in part by DARPA under
Grant No. HR0011-09-1-0015 and by the National Science Foundation
under Grant No. PHY05-51164. This work was completed while T.O. and D.T. were
attending the ``Fundamental Aspects of Superstring Theory'' workshop
at the Kavli Institute for Theoretical Physics at UC Santa Barbara.
They thank the KITP for hospitality. D.T. is grateful to the Theory
Group at CERN and to Perimeter Institute for nice hospitality during
the completion of this work. Research at Perimeter Institute is
supported in part by the Government of Canada through NSERC and by
the Province of Ontario through MRI. J.G. also acknowledges further
support from an NSERC Discovery Grant. The research of T.O. was
supported in part by the National Science Foundation under Grant No.
PHY05-51164. D.T. is partly supported by the NSF grant PHY-05-51164
and by the Department of Energy under Contract DE-FG02-91ER40618.


\appendix
\section{Weyl transforms between  metrics}
\label{Weyl}

In this appendix we discuss the two Weyl transformations relating
$\R^4$ and $AdS_2\times S^2$, which we have used in Section
\ref{sec-thooft}. The first transformation is relevant for the
circular 't Hooft loop computation, and the second one for the
straight line.

Let us parametrize $\R^4$ using two sets of polar coordinates so
that
\be
ds^2_{\mathbb{R}^4}=dl^2+l^2d\psi^2+dL^2+L^2 d\phi^2.
\ee
These coordinates are relevant for a circular loop, which we take to
be located at $l=a$ and $L=0$. By making the following change of
coordinates
\be
\Omega^2={(l^2+L^2-a^2)^2+4a^2L^2\over 4a^2}
={a^2\over(\cosh\rho-\cos\theta)^2}\,,\,\,\,
l=\Omega\sinh\rho\,,\,\,\, L=\Omega\sin\theta\,,
\label{ads2s2}
\ee
we find that the metric becomes
\be
ds^2_{\mathbb{R}^4}=\Omega^2\left(ds^2_{AdS_2}
+d\theta^2+\sin^2\theta\,d\phi^2\right) \,,
\label{conff}
\ee
where
\ba
ds^2_{AdS_2}=d\rho^2+\sinh^2\rho\, d\psi^2
\ea
is the metric on the $AdS_2$ Poincar\'e disk  in global coordinates.
Thus $\R^4$ is conformal to $AdS_2\times S^2$. Note that the loop,
which was located at $l=a,\, L=0$ in $\R^4$, gets mapped to the
conformal boundary  of the Poincar\'e disk.

The metric for $\R^4$ can also be written as
\ba ds^2_{\mathbb{R}^4}=dt^2+dr^2+r^2 d\Omega^2_2, \ea
where $d\Omega^2_2$ is the $S^2$ metric. We place the straight line
at $r=0$. In this case the Weyl transformation to $AdS_2\times S^2$
produces  the hyperbolic metric on the upper half plane
\ba
ds^2_{\mathbb{R}^4}=r^2 (ds^{\prime 2}_{AdS_2}+d\Omega^2_2)\, ,
\ea
where
\ba
ds^{\prime 2}_{AdS_2}=\f{dt^2+dr^2}{r^2} \label{UHP}
\ea
is the $AdS_2$ metric in Poincar\'e coordinates.


\section{Cancellation of non-zero modes}
\label{cancellation}

In this appendix, we show the cancellation of the one loop
determinants in (\ref{oneloop}). This will be first done for the
straight line and $\theta=0$ using self-duality of the background,
and then we will generalize it to non-zero $\theta$ and the circular
loop.

Let us package the four-dimensional bosonic fields
into the ten-dimensional gauge field in the order
\ba
(A_M)\equiv (A_1,\ldots, A_3, \phi^1,\ldots, \phi^6, A_0),~~M=1,2,\ldots, 10.
\ea
Without loss of generality we can take $\phi=\phi^1=A_4$ in
(\ref{theta-background}).
Then the non-zero components of the
ten-dimensional background field strength are given by
\ba
F_{ij}=\f{B}{2r^3}\epsilon_{ijk} x^k,~~~
F_{4i}=\f{B}{2r^3}x^i,~~~~~i,j=1,2,3.
\ea
If we let the index $\mu$ take values $\mu=1,\ldots, 4$,
the four-dimensional field strength $F_{\mu\nu}$ is anti-self-dual.
To exploit this,
we represent the
ten-dimensional gamma matrices in terms of four- and six-dimensional ones as
\ba
\Gamma^\mu=\gamma^\mu\otimes 1,~~~~\Gamma^m=\gamma\otimes \gamma^m,~~~~(m=5,6,\ldots, 10)
\ea
and further decompose $\gamma^\mu$ as
\ba
\gamma^\mu&=&\left(
\begin{array}{cc}
0& \sigma^\mu \\
\bar\sigma^\mu&0\\
\end{array}
\right),~~~
\sigma^\mu \bar\sigma^\nu+\sigma^\nu \bar\sigma^\mu=2\delta^{\mu\nu}
\ea
using $\sigma^4=i=-\bar\sigma^4,~\sigma^j=\bar\sigma^j$.

We begin with the fermionic determinant in (\ref{oneloop}),
where it is raised to the power $1/4=1/2\times 1/2$ because
the fermion $\psi$ satisfies the ten-dimensional Weyl
and Majorana conditions (see footnote \ref{Majorana}).
It is convenient to compute the square (we omit the subscript 0
hereafter)
\ba
&&(i\Gamma^M D_M)^2
\nn\\
&=& -D^2 1_{32}+\f{i}2\Gamma^{MN} F_{MN}
\nn\\
&=&
 \left(
\begin{array}{cc}
\displaystyle-D^2 1_2+\f{i}2\sigma^{\mu\nu}F_{\mu\nu}&
 \\
&-D^2 1_2
\end{array}
\right)\otimes 1_8,
\ea
where we defined  $\sigma^{\mu\nu}\equiv(\sigma^\mu\bar\sigma^\nu-\sigma^\nu\bar\sigma^\mu)/2$
and used the anti-self-duality of $F_{\mu\nu}$.
Thus the fermionic contribution is
\ba
&&\left[{\rm det}_f\left(i\Gamma^M D_{M}\right)\right]^{1/4}
\nn\\
&=&
\left[\det{}'
(i\Gamma^M D_M)^2
\right]^{1/8}
\nn\\
&=&\det{}'\left(-D^2 1_2+\f{i}2\sigma^{\mu\nu}F_{\mu\nu}\right)\det{}'(-D^2)^2,
\label{fermion-con}
\ea
where the prime indicates the omission of zero modes.

For the bosons, we have
\ba
 \left[{\rm det}_b\left(-\delta^{MN}D^2+2i F^{MN}\right)\right]^{-1/2}=
\det{}'(-D^2)^{-3} \det{}'(
-D^2\delta_{\mu\nu}+2i F_{\mu\nu}
)^{-1/2}.
\ea
Observe that
\ba
&&-D^2\delta_{\mu\nu}+2i F_{\mu\nu}
\nn\\
&=&
\half
\bar\sigma_{\mu}^{~\dot \alpha\alpha}
\left(-D^2 1_2
+\f{i}2\sigma^{\rho\sigma}F_{\rho\sigma}
\right)
\hskip -1mm
{}_{ \alpha}^{~\beta}
\delta_{\dot\alpha}^{~\dot\beta}
\sigma_{\nu \beta\dot \beta}.
\ea
By treating $(\dot\alpha\alpha)$ and $(\beta\dot\beta)$ as single indices
taking four values, we can regard
$\bar\sigma_{\mu}^{~\dot \alpha\alpha}$ and $\sigma_{\nu\beta\dot\beta}$
as $4\times 4$ matrices.
Then we see that
\ba
\det{}'(
-D^2\delta_{\mu\nu}+2 i F_{\mu\nu})=\det{}'\left(
-D^2 1_2+\f{i}2\sigma^{\mu\nu}F_{\mu\nu}\right)^2.
\ea
Thus the bosonic contribution can be written as
\ba
 \left[{\rm det}_b\left(-\delta^{MN}D^2+2i F^{MN}\right)\right]^{-1/2}=
\det{}'(-D^2)^{-3} \det{}'\left(
-D^2+\f{i}2\sigma^{\mu\nu}F_{\mu\nu}\right)^{-1}.\label{boson-con}
\ea

Finally, the ghost contribution is simply given by
\ba
{\rm det}_g(-D^2) =\det{}'(-D^2) \label{ghost-con}.
\ea
We  see that the three contributions
(\ref{fermion-con}), (\ref{boson-con}) and (\ref{ghost-con})
cancel out in (\ref{oneloop}):
\ba
{\left[\det_f\left(i\Gamma^M D_{ M}\right)\right]^{1/4}\det_g\left(-D^2\right)\over \left[\det_b\left(-\delta^{MN}D^2+2i F^{MN}\right)\right]^{1/2}}
=1.
\ea

When the theta angle is turned on,
the background fields change to
\ba
F_{0i}=ig^2\theta\f{B}{16\pi^2} \f{x^i}{r^3},~~
F_{ij}=\f{B}{2r^3}\epsilon_{ijk} x^k,~~~
F_{4i}=\f{B}{2r^3}x^i\left(1+\f{g^4\theta^2}{64\pi^4}\right)^{1/2}
\ea
with other components of $F_{MN}$ vanishing.
If we rotate the gauge field and Gamma matrices into
 \ba
 \left(
\begin{array}{cc}
 A'_0\\
 A'_4
\end{array}
 \right)
\equiv
R(\theta)
 \left(
\begin{array}{cc}
 A_0\\
 A_4
\end{array}
 \right)
,
~~~~~
 \left(
\begin{array}{cc}
\Gamma^{\prime 0}\\
\Gamma'^4
\end{array}
 \right)
\equiv
R(\theta)
 \left(
\begin{array}{cc}
\Gamma^0\\
\Gamma^4
\end{array}
 \right)
 \ea
by a complex orthogonal matrix
\ba
R(\theta)=
 \left(\displaystyle
\begin{array}{cc}
\displaystyle\left(1+\f{g^4\theta^2}{64\pi^4}\right)^{1/2}
\displaystyle&\displaystyle -i\f{g^2\theta}{8\pi^2}\\
\displaystyle i\f{g^2\theta}{8\pi^2}
&\displaystyle \left(1+\f{g^4\theta^2}{64\pi^4}\right)^{1/2}
\end{array}
 \right),
\ea
with $A'_M=A_M,~\Gamma'^M=\Gamma^M$ for other $M$,
the corresponding field strength $F'_{\mu\nu}$ remains anti-self-dual,
and we have the relation
\ba
\Gamma^{MN}F_{MN}={\Gamma'}{}^{\mu\nu}F'_{\mu\nu}.
\label{ASD-theta}
\ea
Thus  the proof for the cancellation of non-zero modes still goes
through for the straight line. The relation (\ref{ASD-theta}) also
implies that the background remains BPS for $\theta\neq 0$.

These arguments for the line can be mapped to the circular loop
formally by a conformal transformation and by using different
gauge fixing terms that are generated by the transformation. Since
such cancellation does not depend on the gauge fixing procedure, the
non-zero modes are also cancelled for the circular loop.


\section{Volumes of  groups and coset spaces}
\label{volume}

To relate the integral in (\ref{W2}) to ${\rm Vol}(H/T)$, let us
consider the special case when $w=0$. The Wilson loop expectation
value is then unity. By substituting (\ref{partition-function}) and
$w=0$ in (\ref{W2}), we obtain
\ba
1= \left( \f{2}{\pi g^2} \right)^{\dim (G)/2} \f{{\rm
Vol}(G/T)}{ |\Wcal|} \int_{\tfrak} [dX]
 e^{-\f{2}{g^2}\langle X, X\rangle}
\prod_{\alpha>0}
\alpha(X)^2\,.\label{VolG}
\ea
This formula is very general and can be applied
to the case when $G$ is replaced by $H$, the stability group of
the highest weight $w$, and is given by
\ba
1=
\left(
\f{2}{\pi g^2}
\right)^{\dim (H)/2}
\f{{\rm Vol}(H/T)}{ |\Wcal(H)|}
\int_{\tfrak} [dX]
 e^{-\f{2}{g^2}\langle X, X\rangle}
\mathop{\prod_{\alpha>0}}_{\langle \alpha, w\rangle=0}
\alpha(X)^2\,. \label{VolH}
\ea
We can use (\ref{VolH}) to eliminate the integral in (\ref{W2})
in favor of ${\rm Vol}(H/T)$.
Though (\ref{VolG}) and (\ref{VolH}) are sufficient for our purposes,
we note that such integrals have been explicitly evaluated
in \cite{1980InMat..56...93M} in terms of the
exponents of the Lie algebras.

Let us now derive the relation (\ref{map2}). First consider setting
$(g^2, w)$ to $(2, B)$. By taking the ratio of (\ref{VolG}) and
(\ref{VolH}), we obtain \ba &&{\rm Vol}(G/H)
\nn\\
&&~~~~~~= \pi^{\dim(G/H)/2}\f{|\Wcal|}{|\Wcal(H)|}
\f{\displaystyle\int_{\mathfrak t} [dX] e^{-\langle X,X\rangle}
\prod_{\alpha>0,\,\alpha(B)=0} \alpha(X)^2 }{
\displaystyle\int_{\mathfrak t} [dX] e^{-\langle X,X\rangle}
\prod_{\alpha>0} \alpha(X)^2 }. \label{G/H} \ea
Next by setting $(g^2, G, H, \Wcal, \Wcal(H), \ldots)$ to
$(2/n_\gfrak, \LG, \LH, \Wcal(\LG), \Wcal(\LH),\ldots)$, we derive
by the same procedure
 \ba
&& {\rm Vol}(\LG/\LL H)
\nn\\
&&
~~~~~=
 \left(\f{\pi}{n_\gfrak}\right)^{
 \dim(\LG/\LL H)/2}\f{|\Wcal(\LG)|}{|\Wcal(\LH)|}
 \f{
\displaystyle
\int_{\LL \mathfrak t}
 [d\LL X] e^{-n_\gfrak \langle \LL X,\LL X\rangle}
\prod_{\LL \alpha>0,~\langle \LL w, \LL \alpha \rangle=0}
\LLalpha(\LL X)^2
}{
\displaystyle
\int_{\LL \mathfrak t} [d\LL X] e^{-n_\gfrak \langle \LL X,\LL X\rangle}\prod_{\LL \alpha>0}
\LLalpha(\LL X)^2}.
 \label{LG/LH}
 \ea
Under the isomorphisms between $\tfrak,\, \tfrak^*,\LLt$ and $\LLt^*$,
we can identify $(\LL X, \LLalpha)$ with $(X,\alpha)$.
This involves rescaling of the metric (see footnote \ref{metrics})
$\langle \LL X,\LL X\rangle= \langle X,X\rangle /n_\gfrak$
and the relation
\ba
\LLalpha(\LL X)= \f{2\alpha(X)}{n_\gfrak\langle \alpha,\alpha\rangle }.
\ea
Noting that $\dim \LG=\dim G,\, \dim \LL H=\dim H,\, \Wcal \simeq \Wcal(\LG),\,
\Wcal(\LH)\simeq \Wcal(H)$,
we can put (\ref{LG/LH}) into the form
\ba
&&{\rm Vol}(\LG/\LL H)
\nn\\
&& = (\pi n_\gfrak)^{\dim(G/ H)/2} \displaystyle
\f{|\Wcal|}{|\Wcal(H)|} \Bigg
(\mathop{\prod_{\alpha>0}}_{\alpha(B)\neq 0} \f{\langle
\alpha,\alpha\rangle}2
\Bigg
)^2
\f{\displaystyle
\int_{ \mathfrak t}
 [d X] e^{-\langle  X, X\rangle}
\prod_{ \alpha>0,\,  \alpha(B)=0}
\alpha( X)^2
}{\displaystyle
\int_{ \mathfrak t} [d X] e^{-\langle X, X\rangle}\prod_{ \alpha>0}
\alpha( X)^2}.
\label{LG/LH2}
\ea
By cancelling the ratios of integrals in (\ref{G/H}) and
(\ref{LG/LH2}), and using the relation
$\langle\LLalpha,\LLw\rangle=2\alpha(B)/\langle \alpha,\alpha\rangle
n_\gfrak$, we finally obtain (\ref{map2}).


\section{Examples of Wilson loop expectation values}
\label{ex-wil}

To illustrate the formula (\ref{leading-mixed2-1}) for the Wilson loop
and compare it with the corresponding formula (\ref{hooftfinal})
for the 't Hooft loop, let us give some examples.

 \noindent $\bullet$ $G=SU(2)$ and $SO(3)$.
Irreducible representations are labeled by the spin $j$. According
to the formula (\ref{leading-mixed2-1}) the expectation value at
large $g^2$ is
\ba
\langle W(j)\rangle_{G,\tau}= \exp\left(\f{j^2}4 g^2\right)j^2 g^2.
\label{wil-j} \ea

\noindent $\bullet$ $G=U(N)$.
A highest weight $w=[m_1,\ldots, m_N]$
satisfies $m_1\geq \ldots \geq m_N$.

    $\vartriangleright w=[k,0,\ldots, 0]$. For the rank-$k$ symmetric representation
\ba
 \langle W\left([k,0,\ldots, 0]\right)\rangle_{G,\tau}
= \exp\left({k^2 \over 8} g^2\right) \left({k^2 g^2 \over 4}\right)^{N-1}
{1\over (N-1)!} \,. \label{wil-sym}
 \ea

    $\vartriangleright w=
[\hskip 1pt \displaystyle\mathop{\underbrace{1,\ldots,1}}_{k\, {\rm times}}
,0,\ldots,0]$.
The rank-$k$ antisymmetric representation gives
\vspace{-4mm}
\ba
 \langle W\left([1,\ldots,1,0,\ldots,0]\right)\rangle_{G,\tau} = \exp\left({k \over 8} g^2\right)
 \left({g^2\over 4}\right)^{k(N-k)}  {\prod_{n=1}^{k-1} n!\over \prod_{n=1}^{k} (N-n)!}\,.
 \label{wil-antisym}
 \ea

   $\vartriangleright w=[m_1,m_2,\ldots,m_N]$ with $m_1>m_2\ldots > m_N$.
For such a representation
    \ba
 \langle W
\left([m_1,m_2,\ldots,m_N]\right) \rangle_{G,\tau} =\exp\left(\f{\sum_i m_i^2 }{8}g^2 \right)
\left({g^2\over 4}\right)^{ \f{N(N-1)}2}
 \f{\prod_{i<j}(m_i-m_j)^2}{\prod_{n=1}^{N-1} n!}.
\label{wil-gen}
 \ea
The results (\ref{wil-j})-(\ref{wil-gen}) agree with
(\ref{thooft-j})-(\ref{thooft-gen})
via the $S$-duality map $g^2\ra \LLgc^2=g^2|\tau|^2$.


\subsection{The method of orthogonal polynomials}

In Section \ref{sec-wilson} we have derived an expression for the
expectation value of a Wilson loop $W(R)$ at strong coupling
(\ref{leading-mixed2-1}) that is valid for an arbitrary gauge group
$G$. For illustration purposes, we present here an alternative
derivation of $\langle W(R)\rangle$ for the case $G=U(N)$ obtained
with the method of orthogonal polynomials.

As we saw in Section \ref{sec-wilson}, the Wilson loop $W(R)$ can be
written as a sum over weights of the representation $R$. At large
$g$, the terms that dominate are the ones corresponding to the
weights obtained from the highest weight $w$ by the action of the
Weyl group. For $U(N)$, the highest weight can be labeled as
$w=(m_1,\ldots ,m_N)$, where the integers $m_i$ are ordered:
$m_1\geq m_2\geq \ldots\geq m_N$. We introduce integers $N_I$
($I=1,\ldots, M)$ such that $w=(m_1,\ldots, m_N)$ contains $M$
distinct integers, with the $I$-th one appearing $N_I$ times. One
can rotate the matrix $M$ in (\ref{M-integral}) to a diagonal
configuration with eigenvalues $\{x_i\}$ at the cost of introducing
a  Vandermonde determinant $\Delta^2=\prod_{i<j}(x_i-x_j)^2$  in the
integration measure. This determinant can be rewritten in terms of
polynomials that are orthogonal with respect to the Gaussian
measure, so that the integral to compute becomes
\bea \langle W(R) \rangle \simeq \oo{\prod_{I=1}^M N_I!} \int
\Big(\prod_{i=1}^N dx_i\Big) \det(\{P_{j-1}(x_i)\})^2\,e^{-\sum_i
x_i^2}\, e^{\frac{g}{\sqrt 2}\sum_i m_i x_i}\,. \nonumber \eea
These polynomials are normalized Hermite polynomials:
$P_n(x)\equiv H_n(x)/\sqrt{2^n n! \sqrt \pi}$ \cite{Drukker:2000rr}.
Completing the squares in the exponentials one readily finds
\bea
\langle W(R)\rangle\simeq \oo{\prod_{I=1}^M N_I!}
 e^{\frac{g^2}{8}\sum_i m^2_i}\int \Big(\prod_{i=1}^N dx_i\Big)
 \det(\{P_{j-1}(x_i)\})^2\,e^{-\sum_i
 \left(x_i-m_i\frac{g}{2\sqrt 2}\right)^2}\,.
\label{U(N)a}
\eea
To obtain the polynomial corrections to the exponential behavior, we
need
\bea {\cal I}^{(k)}_{ij}\equiv \int_{-\infty}^\infty dx\, e^{-
\left(x-k\frac{g}{2\sqrt 2}\right)^2} P_i(x)P_j(x)\,. \nonumber \eea
For this, it is useful to use the contour integral representation of
the Hermite polynomials
\bea H_n(x)=\frac{n!}{2\pi i}\oint_{\cal C} dt \, \frac{e^{-t^2+2 t
x}}{t^{n+1}}\,,\nonumber \eea
where ${\cal C}$ encircles the origin counterclockwise. We find
\bea {\cal I}^{(k)}_{ij}=\frac{1}{\sqrt{i!j!}}\left(\frac{k\,
g}{2}\right)^{i+j}\sum_{\ell=0}^{{\rm min}(i,j)} \ell!
\left(\begin{array}{c}i\\
\ell\end{array}\right)\left(\begin{array}{c}j\\
\ell\end{array}\right)\left(\frac{2}{k\,g}\right)^{2\ell}\,,
\nonumber \eea
which can be expressed in terms of a confluent hypergeometric
function of the second kind $U(a,b,z)$ as
\bea {\cal I}^{(k)}_{ij}=\frac{(-1)^i}{\sqrt{i!j!}}\left(\frac{k\,
g}{2}\right)^{j-i}
U\left(-i,1-i+j,-\frac{k^2g^2}{4}\right)\,.\label{I} \eea
By applying this to (\ref{U(N)a}), we arrive to
\bea
\langle W(R)  \rangle&\simeq& \f{(-1)^{\frac{N(N-1)}{2}}}
{\prod_{I=1}^M N_I!}
 \frac{e^{\frac{g^2}{8}\sum_i m^2_i}}{\prod_{n=0}^{N-1}n!}
 \sum_{\sigma,\sigma'\in S_N}
 \mbox{sign}(\sigma)\,\mbox{sign}(\sigma')
 \cr && \hskip 1cm \times \prod_{i=0}^{N-1}
m_{i+1}^{\sigma'(i)-\sigma(i)}\,U\left(-\sigma(i),1-\sigma(i)
+\sigma'(i),-\frac{m_{i+1}^2g^2}{4}\right)\,,
\label{Wfinal}
\eea
where $\sigma$ and $\sigma'$ permute $\{0,1,\ldots, N-1\}$.

We can now specialize (\ref{Wfinal}) to the representations of
$G=U(N)$ that we have considered above, see equations
(\ref{wil-sym})-(\ref{wil-gen}).

$\vartriangleright w=[k,0,\ldots, 0]$. For the rank-$k$ symmetric
representation, we need to use the following limits of the confluent
hypergeometric function
\bea \lim_{x\to 0} x^{j-i} \, U(-i,1-i+j,- x^2) = (-1)^i \, i!
\,\delta_{ij} \,, \label{limitU1} \eea
{\it i.e.}, ${\cal I}^{(0)}_{ij} = \delta_{ij}$, and
\bea \lim_{x\to \infty} U(-i,1-i+j,-x)= (-1)^i\, x^i
\label{limit}\,. \eea
We get
\bea \langle W(R)\rangle &\simeq& \frac{e^{\frac{g^2}{8}
k^2}}{(N-1)!\prod_{n=0}^{N-1}n!} \cr && \hskip .2cm \times
\sum_{\sigma,\sigma'\in S_N}
\mbox{sign}(\sigma)\,\mbox{sign}(\sigma')
\left(\frac{g}{2}\right)^{2\sigma(0)}
k^{\sigma'(0)+\sigma(0)}\prod_{i=1}^{N-1}\sigma(i)!\,
\delta_{\sigma(i)\sigma'(i)}\,. \nonumber \eea
Since we are interested in the highest power of $g$ we must select
$\sigma(0)=N-1$. The product over the Kronecker deltas imposes that
$\sigma=\sigma'$, while the product over the factorials gives
$\prod_{n=0}^{N-2}n!$. By summing over $(N-1)!$ permutations, we
have
\bea \langle W(R) \rangle \simeq \oo{(N-1)!} e^{\frac{g^2}{8} k^2}
\left(\frac{g}{2}\right)^{2(N-1)}k^{2(N-1)}\,.\nonumber \eea
This is what we found in (\ref{wil-sym}).

$\vartriangleright w= [\hskip 1pt
\displaystyle\mathop{\underbrace{1,\ldots,1}}_{k\, {\rm times}}
,0,\ldots,0]$. To compute the rank-$k$ antisymmetric representation
we use again the limit (\ref{limitU1}) but, because of the
degeneracy of the non-zero $m_i$'s, we need in this case the full
expression (\ref{I}) for the confluent hypergeometric function,
since some of the leading polynomial corrections cancel for
combinatorial reasons. We find
\bea \langle W(R)\rangle &\simeq& \frac{e^{\frac{g^2}{8}
k}}{k!(N-k)!\prod_{n=0}^{N-1}n!} \sum_{\sigma,\sigma'\in S_N}
\mbox{sign}(\sigma)\,\mbox{sign}(\sigma')
\left(\prod_{j=k}^{N-1}\sigma(j)!\,\delta_{\sigma(j)\sigma'(j)}\right)
 \cr && \hskip .2cm \times \left(\prod_{i=0}^{k-1}
\sigma(i)!\,\sigma'(i)!\sum_{\ell=0}^{{\rm
min}(\sigma(i),\sigma'(i))}
\frac{1}{\ell!(\sigma(i)-\ell)!(\sigma'(i)-\ell)!}
\left(\frac{g^2}{4}\right)^{\sigma(i)-\ell} \right) \,. \nonumber
\eea
To obtain the leading order in the coupling constant we require
$\{\sigma(0),\ldots,\sigma(k-1)\}=\{N-k,\ldots,N-1\}$, from which it
also follows $\{\sigma(k),\ldots,\sigma(N-1)\}=\{0,\ldots,N-k-1\}$
and, because of the Kronecker deltas,
$\{\sigma'(0),\ldots,\sigma'(k-1)\}=\{N-k,\ldots,N-1\}$ and
$\{\sigma'(k),\ldots,\sigma'(N-1)\}=\{0,\ldots,N-k-1\}$. The power
of the coupling constant $g^2/4$, without considering the subleading
contributions coming from the sums over $\ell$'s, is then given by
$\sum_{n=N-k}^{N-1}=k(2N-k-1)/2$. Because of the degeneracy of the
$m_i$'s there are cancellations though and this power is reduced of
$\sum_{n=0}^{k-1}n=k(k-1)/2$ to give $(g^2/4)^{k(N-k)}$. Selecting
only the terms in the sums over $\ell$'s that produce this power and
performing the sums over $k!$ and $(N-k)!$ permutations, one finally
finds the same result as (\ref{wil-antisym}).

 $\vartriangleright w=[m_1,m_2,\ldots,m_N]$ with $m_1>m_2\ldots > m_N$.
In this case we simply expand the hypergeometric functions  for
large $g$ using (\ref{limit}). It is easy to see that (\ref{Wfinal})
becomes
\bea \langle W(R) \rangle&\simeq& \frac{e^{\frac{g^2}{8}\sum_i
m^2_i}}{\prod_{n=0}^{N-1}n!}\left(\frac{g}{2}\right)^{N(N-1)}
\sum_{\sigma,\sigma'\in S_N}
\mbox{sign}(\sigma)\,\mbox{sign}(\sigma')
\prod_{i=0}^{N-1}m^{\sigma(i)+\sigma'(i)}_{i+1}\cr &=&
\frac{e^{\frac{g^2}{8} \sum_i
m^2_i}}{\prod_{n=0}^{N-1}n!}\left(\frac{g}{2}\right)^{N(N-1)}
\prod_{i<j=1}^N(m_i-m_j)^2\,,\nonumber \eea
in agreement with (\ref{wil-gen}).

\vfill\eject
\bibliography{thooft}

\end{document}